\documentclass{elsarticle}
\usepackage{fullpage}

\usepackage[utf8]{inputenc}
\usepackage{graphicx}
\usepackage{bm}
\usepackage[hidelinks]{hyperref}
\usepackage{amsmath,amssymb}
\usepackage{mathtools}
\usepackage[version=4]{mhchem}
\usepackage{siunitx}
\usepackage{longtable,tabularx}
\usepackage{algorithm}
\usepackage{algpseudocode}
\usepackage{booktabs}
\usepackage{multirow}
\usepackage[textsize=small]{todonotes}

\newcommand{\bu}{\mathbf{u}}
\newcommand{\bq}{\mathbf{q}}
\newcommand{\bx}{\mathbf{x}}

\newcommand{\bU}{\mathbf{U}}
\newcommand{\bF}{\mathbf{F}}
\newcommand{\Kn}{\mathrm{Kn}}

\journal{arXiv}

\begin{document}

\begin{frontmatter}

\title{Physics-Based Machine Learning Closures and Wall Models for Hypersonic Transition–Continuum Boundary Layer Predictions}

\author[ND]{Ashish S. Nair}
\author[TAMU]{Narendra Singh}
\author[UCI]{Marco Panesi}
\author[Oxford]{Justin Sirignano}
\author[ND]{Jonathan F. MacArt\corref{cor1}}

\affiliation[ND]{
  organization={Department of Aerospace and Mechanical Engineering, University of Notre Dame},
  city={Notre Dame},
  state={IN},
  postcode={46556},
  country={USA}
}%

\affiliation[TAMU]{
  organization={Department of Aerospace Engineering, Texas A\&M University},
  city={College Station},
  state={TX},
  postcode={77840},
  country={USA}
}%

\affiliation[UCI]{
  organization={Department of Mechanical and Aerospace Engineering, University of California, Irvine},
  city={Irvine},
  state={CA},
  postcode={92697},
  country={USA}
}%

\affiliation[Oxford]{organization={Mathematical Institute, University of Oxford},
            city={Oxford},
            postcode={OX2 6GG}, 
            country={UK}}

\cortext[cor1]{Corresponding author. \texttt{jmacart@nd.edu}}

\begin{abstract}

Modeling rarefied hypersonic flows remains a fundamental challenge due to the breakdown of classical continuum assumptions in the transition-continuum regime, where the Knudsen number ranges from approximately 0.1 to 10. Conventional Navier--Stokes--Fourier (NSF) models with empirical slip-wall boundary conditions fail to accurately predict nonequilibrium effects such as velocity slip, temperature jump, and shock structure deviations. In this work, we develop a physics-constrained machine learning framework that augments bulk transport models and boundary conditions to extend the applicability of continuum solvers in nonequilibrium hypersonic regimes. We employ deep learning PDE models (DPMs) for the viscous stress and heat flux embedded in the governing equations and trained via adjoint-based optimization. Three closure strategies are evaluated for two-dimensional supersonic flat-plate flows in argon across a range of Mach and Knudsen numbers. Additionally, we introduce a novel wall model based on a {mixture of skewed Gaussian approximations} of the particle velocity distribution function. This wall model replaces empirical slip conditions with physically informed, data-driven boundary conditions for the streamwise velocity and wall temperature. Our results show that a trace-free anisotropic viscosity model,  paired with the skewed-Gaussian distribution function wall model, achieves significantly improved accuracy in predicting both bulk and wall quantities, particularly at high-Mach and high-Knudsen number regimes. Strategies such as parallel training across multiple Knudsen numbers and inclusion of high-Mach data during training are shown to enhance model generalization. Increasing model complexity yields diminishing returns for out-of-sample cases, underscoring the need to balance {degrees of freedom} and overfitting. To further assess generalization, we test the best-performing model on a geometrically out-of-sample wedge with increasing wall angles, for which the accuracy of the augmented predictions is consistent with the deviation from the training configuration. This work establishes data-driven, physics-consistent strategies for improving hypersonic flow modeling for regimes in which conventional continuum approaches are invalid. 

\end{abstract}

\end{frontmatter}

\section{Introduction} \label{sec:Intro}

The accurate design and development of hypersonic vehicles rely on flow models that can handle transition-continuum flow regimes with Knudsen numbers Kn ranging from approximately 0.1 to 10. Currently available computationally efficient hydrodynamic models are not reliable in these flow regimes due to their reliance on continuum assumptions. Traditional computational fluid dynamics (CFD) breaks down as rarefaction effects (finite molecular mean free paths) lead to velocity slip and temperature jump at wall boundaries. For example, continuum models with no-slip boundaries fail to capture near-wall Knudsen layers, and even with slip/jump corrections they are only reliable for $\mathrm{Kn}\lesssim0.1$~\cite{JJShuFluidVelocitySlip}. Precise simulations of nonequilibrium flows in more rarefied regimes typically require solving the Boltzmann equation using methods such as direct simulation Monte Carlo (DSMC).

The computation of the shock structure at high Mach numbers is particularly interesting in hypersonic flows, especially when additional physical phenomena such as chemical reactions are involved. The shock structure predicted by the Navier--Stokes--Fourier (NSF) equations has been found to be inaccurate even at moderate Mach numbers~\cite{Vincenti1917, Greenshields2007}. Experimental measurements have shown that the shock layer thickness predicted by the NSF equations is too small compared to actual observations, particularly at intermediate Mach numbers~\cite{LiznerAndHornig, Camac, Schmidt1969, Garen1974, Alsemeyer1976}. The failure of the NSF equations to capture shock structure accurately at higher Mach numbers has been attributed to the breakdown of the linear Newton's and Fourier's laws~\cite{Paolucci2018}. To address this, numerous higher-order continuum-fluid closures have been proposed, many derived from the Boltzmann equation and often limited to monatomic gases. These include classical Chapman--Enskog expansions~\cite{Chapman1970}, Grad's 13-moment equations~\cite{Grad1949, Grad1952}, and various regularized or modified Burnett-type models~\cite{Agarwal2001, Jin2001, RegBurett2003, Woods1993}. However, such models often suffer from numerical and physical instabilities and may violate thermodynamic principles. Conversely, particle-based approaches such as DSMC~\cite{Bird1963, Bird1970} and molecular dynamics (MD) simulations can capture detailed nonequilibrium effects and show good agreement with experiments, but their computational cost can be prohibitive in the continuum and transition-continuum regimes. This poses a critical challenge in extending continuum flow solvers to capture nonequilibrium effects without the cost of full Boltzmann solutions.

Boundary conditions, particularly at walls, play a critical role in accurate predictions of hypersonic flows. In many continuum solvers, the empirical Maxwell slip and temperature-jump boundary conditions are employed at walls~\cite{lofthouse2008velocity, Sone2007, Hadjiconstantinou2024MolecularMO}. These first-order slip conditions assume linearity in the Knudsen number and are derived based on simplified kinetic theory. They have been widely used due to their compatibility with existing CFD frameworks and their modest computational overhead. However, these boundary conditions lose accuracy at transition-continuum Knudsen numbers, particularly in flows involving strong rarefaction, surface curvature, or separation~\cite{Lockerby2004, Hadjiconstantinou2003}. This can result in inaccurate predictions of wall shear stress, heat flux, and boundary layer profiles, which are critical for hypersonic vehicle design.

Data-driven methods have seen significant adoption in fluid dynamics, particularly for incompressible flows. In turbulence modeling, these methods have been employed to develop enhanced Reynolds-averaged Navier--Stokes (RANS) and large eddy simulation (LES) closure models~\cite{Ling_Kurzawski_Templeton_2016}. Comprehensive reviews highlight the ability of physics-guided machine learning (ML) turbulence models to reduce model-form uncertainties and improve predictive accuracy~\cite{duraisamy2019turbulence, duraisamy2021perspectives}. Beyond turbulence closures, data-driven reduced-order modeling approaches such as proper orthogonal decomposition (POD) and dynamic mode decomposition (DMD) have been applied to tasks including system identification and autonomous control~\cite{Brunton_2020}. A particularly promising development is the emergence of physics-informed neural networks (PINNs), which embed partial differential equation (PDE) residuals into the training process~\cite{RAISSI2019686, KarniadakisPINNS}. PINNs have been effective for various forward and inverse problems in incompressible flow regimes. These efforts collectively demonstrate how integrating data-driven models with physics-based constraints can enrich traditional fluid flow simulations. More recently, attention has turned toward extending these techniques to compressible flows. While still limited in number, early studies have shown promise despite the added complexity introduced by shock waves, compressibility effects, and the tight coupling of hydrodynamics and thermodynamics. Neural networks have been explored for tasks such as shock detection and compressible turbulence modeling, including PINN-based approaches applied to supersonic flows with shocks~\cite{cai2021physicsinformedneuralnetworkspinns}. Although most ML-based turbulence studies remain focused on low Mach number flows, the growing body of work in compressible regimes suggests strong potential. With appropriate incorporation of physical constraints, ML models are increasingly capable of addressing challenges in compressible flow, from closure modeling to rapid flow prediction.

Extending \emph{a priori} data-driven techniques to high-speed and hypersonic regimes is an active area of research. Hypersonic flows involve complex nonequilibrium phenomena such as thermal and chemical transport, ionization, and rarefaction at high altitudes. These effects introduce substantial challenges to modeling efforts. Recent efforts have explored neural network–based closures and surrogate models to capture these complexities. One approach employs deep operator networks (``DeepONet'') to model coupled flow and finite-rate chemistry in a normal shock~\cite{Mao_2021}.
Despite these advancements, a major limitation remains: conventional black-box neural networks often lack adherence to governing physics. This disconnect can result in models that fit observed data yet violate conservation laws or exhibit unstable behavior when applied to new conditions. The issue is especially pronounced in hypersonics, where available data are limited and operational regimes often extend beyond training conditions. Researchers have shown that without embedded physical constraints, ML-based turbulence closures and surrogate models may produce nonphysical outputs or fail to generalize across flow configurations~\cite{duraisamy2021perspectives}. These challenges highlight the importance of incorporating physical priors or constraints into ML training~\cite{sirignano2020dpm}. In summary, while \emph{a priori} ML models can achieve strong empirical fits, integrating physical constraints is critical for consistent and trustworthy modeling in hypersonics applications~\cite{Zanardi_2023}.

To address these challenges, we employ the deep learning PDE model (DPM) approach \cite{sirignano2020dpm}, which integrates the training of neural networks with the solution of the governing equations, ensuring consistency between the learned model and the physical system. This is achieved by solving the adjoints of the governing equations to provide the gradients needed for neural network parameter optimization. The adjoint equations are constructed using automatic differentiation (AD) over the fully discretized flow solver. This methodology has been successfully applied to turbulence modeling in LES~\cite{sirignano2020dpm, macart2021embedded, sirignano2023bluff} and RANS~\cite{sirignano2021pde, kakka2025neuralnetworkaugmentededdyviscosity} as well as 1D transition-continuum shock predictions~\cite{nair2023deep, kryger2024optimizationsecondordertransportmodels}. In this work, we extend the DPM framework to predictions of two-dimensional (2D) hypersonic boundary layer flows including improved wall models for rarefied hypersonic flow conditions.

In hypersonic flow modeling, accurate treatment of gas–surface interactions is critical, particularly near walls where the continuum assumption breaks down. Traditional wall boundary conditions in continuum solvers rely on the Maxwell model~\cite{maxwell1879vii}, which prescribes the slip velocity and temperature jump based on first-order corrections from kinetic theory. These are typically implemented using empirical formulations involving accommodation coefficients~\cite{Sone2007, Sharipov2011} and are known to lose accuracy in the transition-continuum regime where the velocity distribution deviates significantly from equilibrium. Extensions using Cercignani--Lampis and higher-order velocity-slip/temperature-jump models offer partial improvement but rely nonetheless on near-Maxwellian assumptions~\cite{Hadjiconstantinou2003, Lockerby2004}.

We additionally introduce a class of data-driven wall models based on reconstructed particle velocity distribution functions rather than prescribed macroscopic quantities. These models represent the velocity distribution as a convex combination of skewed Gaussian functions, inspired by formulations which have successfully captured bimodal and otherwise nonequilibrium velocity distributions in rarefied flows~\cite{PhysRevE.81.056314, GRAUR201187}. This formulation provides a flexible yet physically grounded basis for capturing the non-Maxwellian character of wall-proximate distributions in hypersonic flows. The macroscopic wall quantities are obtained by analytic integration of the modeled distribution functions. This approach replaces empirical wall closure models with a physically informed alternative derived from kinetic theory, thereby addressing one of the key limitations of conventional CFD treatments in the transition-continuum regime and extending the validity of continuum solvers to more rarefied conditions.
 
This paper is organized as follows. Section~\ref{sec:Governing Equations} reviews the governing equations and the baseline transport models. Section~\ref{sec:ML Closure} introduces and validates the deep learning framework, transport closure models, and Boltzmann distribution function-based wall models. Section~\ref{sec:Results} presents the application of the optimized transport models and wall models to two-dimensional boundary layer flows. Section~\ref{sec:out-of-sample-geo} tests the generalization of flat plate-trained models to an unseen-in-training geometry. Finally, Section~\ref{sec:conclusion} summarizes the key findings and outlines directions for future work.

\section{Governing Equations, Classical Closures, and Numerical Methods} \label{sec:Governing Equations}

We consider applications to steady-state hypersonic flows over a flat plate in argon, initialized with freestream pressure and temperature $p_{\infty} = 6.667$~Pa and $T_{\infty} = 300$~K. The gas dynamics are governed by the compressible Navier--Stokes equations, which describe the conservation of mass, momentum, and energy,
\begin{equation} \label{eq:1D N-s}
\frac{\partial \mathbf{U}}{\partial t} + \nabla\cdot ( \mathbf{F}_{d} - \mathbf{F}_{c} ) =0,
\end{equation}
where the conserved variables, convective fluxes, and diffusive fluxes are given by
\begin{equation}
\mathbf{U}=\left[\begin{array}{c}
\rho \\
\rho \bu \\
\rho E
\end{array}\right], \quad \mathbf{F}_{c}=\left[\begin{array}{c}
\rho \bu \\
\rho \bu\otimes\bu + p\mathbf{I} \\
\rho \bu H
\end{array}\right], \quad \mathbf{F}_{d}=\left[\begin{array}{c}
0 \\
\tau(\bU;\theta) \\
\tau(\bU;\theta) \cdot \bu - \bq(\bU;\theta)
\end{array}\right].
\end{equation}
Here, $\rho$ is the density, $\bu\in\mathbb{R}^d$ is the $d$-dimensional velocity vector, $p$ is the pressure, $E = e + \bu^\top\bu/2$ is the total energy, where $e$ is the internal energy per unit mass, and  $\mathbf{I}\in\mathbb{R}^{d\times d}$ is the identity matrix. Total enthalpy is defined as $H = e + p/\rho + \bu^\top\bu/2$. The viscous stress tensor $\tau(\bU;\theta)$ and heat flux vector $\bq(\bU;\theta)$ may be either classical (Newton--Fourier) or ML-augmented closures, depending on the model configuration (see Section~\ref{sec:ML Closure}).

\subsection{Classical Constitutive Closures}

In the classical NSF formulation, the constitutive models for the viscous stress and heat flux are given by
\begin{equation}
  \tau = \mu \left( \nabla \bu + \nabla \bu^\top \right) + \left( \mu_b - \frac{2}{3} \mu \right)(\nabla \cdot \bu)\mathbf{I}, \qquad \bq = -k \nabla T,
  \label{eq:Newton}
\end{equation}
where $\mu$ is the dynamic viscosity, $\mu_b$ is the bulk viscosity, and $k$ is the thermal conductivity. These transport coefficients are temperature-dependent and are typically modeled using power-law relations derived from kinetic theory.
For monatomic gases such as argon, the temperature dependence of viscosity is modeled using 
\begin{equation}
  \mu(T) = \mu_{\text{ref}} \left( \frac{T}{T_{\text{ref}}} \right)^\omega,
  \label{eq:sutherland}
\end{equation}
where $\mu_{\text{ref}}$ is the reference viscosity at $T_{\text{ref}}=300$ K,
and $\omega$ is the viscosity exponent. For argon, $\omega \approx 0.72$ under hard-sphere molecular assumptions, although values between 0.68 and 0.75 are used in practice depending on empirical tuning or interaction potentials~\cite{Greenshields2007, Bird1970}.
The thermal conductivity is typically modeled using a temperature exponent, often with the same $\omega$ for monatomic gases, or using an assumed constant Prandtl number, typically taken to be $\text{Pr} = 2/3$ for argon.

\subsection{Slip-Wall Boundary Conditions} \label{sec:slipwall}

For hypersonic boundary layer simulations in rarefied regimes, the classical no-slip and isothermal wall assumptions break down due to the comparable molecular mean free path and characteristic flow scales. In such cases, gas-surface interactions are no longer accurately captured by continuum assumptions, necessitating more detailed wall models that account for nonequilibrium effects~\cite{Lockerby2004}.
Instead, first-order velocity-slip and temperature-jump boundary conditions are typically applied to account for rarefaction effects at the surface~\cite{lofthouse2008velocity}. These are derived from  kinetic theory and provide practical corrections for transition-continuum flows.

The wall slip velocity is modeled as \cite{lofthouse2008velocity} 
\begin{equation}
  V_s = A \left( \frac{2 - \sigma}{\sigma} \right) \Lambda_{HS} \left.\frac{\partial u}{\partial y}\right|_{w}, \label{eq: Slip-velocity}
\end{equation}
where {${\partial u}/{\partial y} |_{w}$} is the wall-normal velocity gradient, $\sigma$ is the tangential momentum accommodation coefficient (taken as $\sigma = 1$ for full accommodation), $A = \sqrt{\pi / 2}$ is a scaling constant, and $\Lambda_{HS}$ is the local hard-sphere mean free path, 
\begin{equation} \label{Eq: Mean-free path} \Lambda_{HS} = \frac{\mu}{\rho} \sqrt{\frac{\pi}{2 R T}}. \end{equation}
The thermal boundary condition accounts for temperature jump at the wall,
\begin{equation} T_0 - T_w = \left( \frac{2 - \alpha}{\alpha} \right) \Lambda_{HS}^T \left.\frac{\partial T}{\partial y}\right|_{w}, \label{eq: Temp-jump} \end{equation}
where $T_0$ is the static temperature of the gas immediately adjacent to the wall, $T_w$ is the wall temperature, which we fix at 300~K, $\alpha = 1$ is the thermal accommodation coefficient, and $\Lambda_{HS}^T$ is the thermal mean free path,
\begin{equation} \Lambda_{HS}^T = \frac{2}{\gamma + 1} \frac{k}{\rho C_v} \sqrt{\frac{\pi}{2 R T}}, \end{equation}
where $R$ is the universal gas constant, $k$ is the thermal conductivity, and $C_v$ is the specific heat at constant volume.
While Maxwell slip and jump boundary conditions improve upon the classical no-slip and isothermal assumptions, they remain semi-empirical and are known to exhibit limited accuracy in the slip and transition-continuum regimes. Specifically, their reliance on accommodation coefficients tuned to specific gas-surface interactions and flow conditions reduces their predictive robustness across a wide range of Knudsen numbers~\cite{karniadakis2005microflows,sone2007molecular,GavasaaneEtAll}. Several studies have shown that these models break down as nonequilibrium effects become more pronounced, especially near walls in hypersonic and microscale flows~\cite{boyd1995predicting,sharipov2011data}.
They assume linear behavior of the streamwise velocity and temperature with respect to the local Knudsen number and are derived under assumptions that neglect strong nonequilibrium effects such as velocity and energy distribution anisotropies near the wall. As such, they often fail to accurately predict wall shear stress, heat flux, or velocity slip in the Knudsen layer for high-speed, high-altitude flows.

To address these limitations, in Section~\ref{Sec : Boltzmann wall models} we introduce a data-driven wall model approach based on modeled nonequilibrium particle velocity distributions. This distribution function-based approach replaces the empirical slip and jump expressions with physically informed, ML-predicted wall quantities that better align with DSMC benchmarks at rarefied conditions.

\subsection{Demonstration Configuration}

We consider applications to flow over a two-dimensional flat plate. We solve the Navier--Stokes equations using a standard fourth-order central-difference scheme~\cite{lele1991compact} for the viscous fluxes, a modified Steger--Warming scheme~\cite{MACCORMACK1989135} for  the inviscid fluxes, and a pseudo-time explicit Euler scheme to advance the solution to steady state, with residual convergence used as a stopping criterion.

\begin{figure}
    \centering \includegraphics[width=0.6\textwidth]{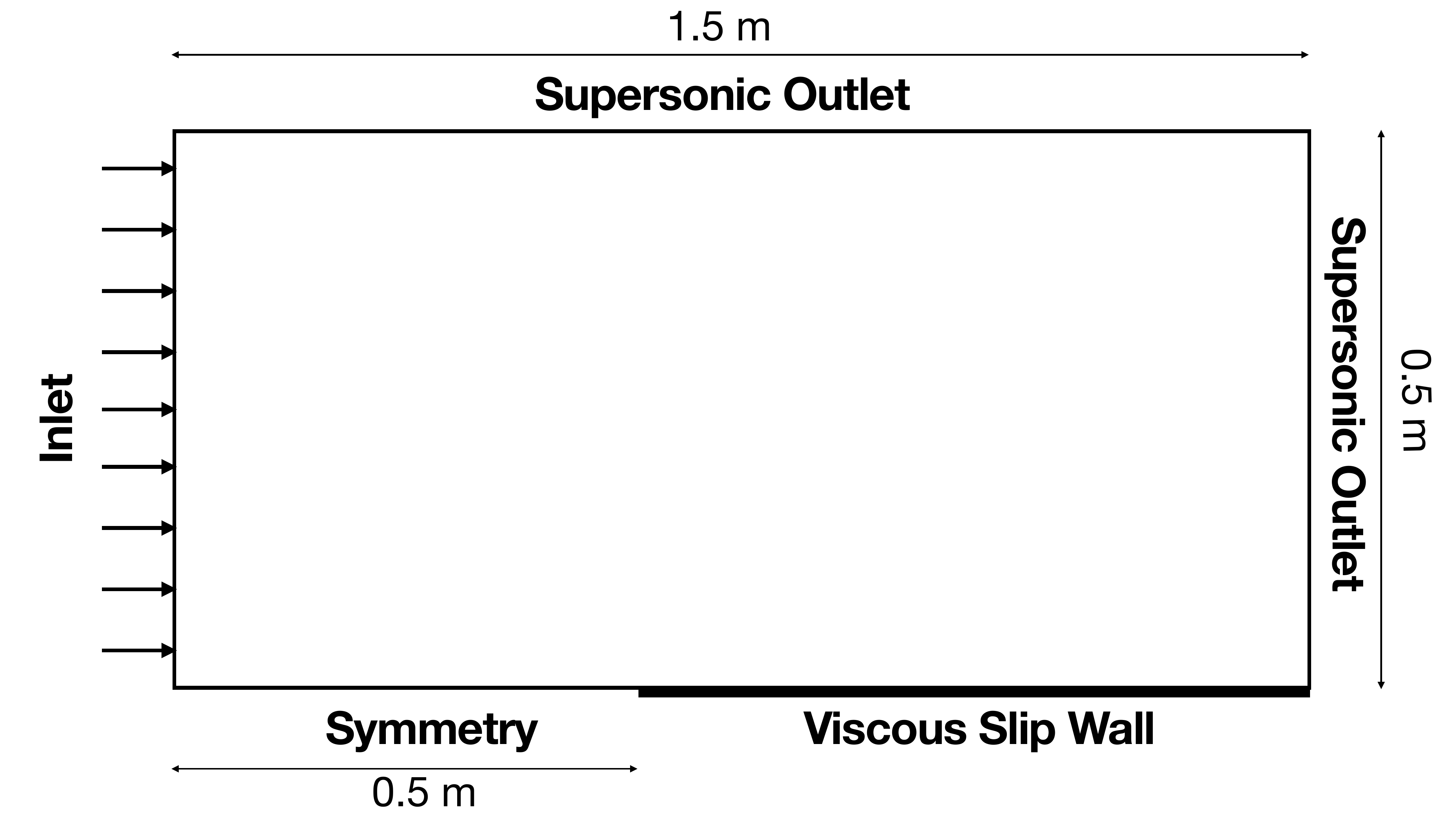} 
    \caption{Schematic of the flat plate geometry showing domain dimensions and boundary conditions.
    } 
    \label{fig:FlatPlateSchem} 
\end{figure}

The physical domain and boundary conditions are illustrated in Figure~\ref{fig:FlatPlateSchem}. The computational mesh consists of 16,384 elements ($N_\xi = 256$, $N_\eta = 256$) spaced nonuniformly with spatial clustering near the leading edge in the $x$-direction and near the wall in the $y$-direction to resolve boundary-layer gradients. The grid stretching uses hyperbolic tangent transformations
\begin{align}
  x_i &= L_x\left(1 - \frac{\tanh(s_x (1 - \xi_i))}{\tanh(s_x)}\right), \quad i \in[0,N_\xi-1], \\
  y_j &= L_y\left(1 - \frac{\tanh(s_y (1 - \eta_j))}{\tanh(s_y)}\right), \quad j \in[0,N_\eta-1],
\end{align}
where $\xi_i=i/(N_\xi-1)$ and $\eta_j=j/(N_\eta-1)$ are the uniformly spaced computational coordinates, $L_x = 1.5$~m and $L_y = 0.5$~m are the domain lengths, and $s_x = s_y = 2$ are the stretching factors.

\section{Machine Learning Closures} \label{sec:ML Closure}

\subsection{Augmented Transport Models} \label{sec:transport_model}

To model deviations from classical continuum behavior, we investigate three machine learning-based augmentation strategies for the Navier--Stokes transport coefficients. These strategies differ in the degree of anisotropy permitted and in their enforcement of physical constraints, such as the trace-free condition for stress tensors in monatomic gases. {In continuum theory, the viscous stress tensor for monatomic gases is inherently trace-free, reflecting the fact that shear stresses should not induce volumetric deformation. Enforcing this property in the augmented models ensures thermodynamic consistency and alignment with the underlying assumptions of kinetic theory, particularly in regimes where continuum descriptions begin to break down.} We investigate the following augmentation frameworks for the transport-coefficient models.
\begin{enumerate}
    \item \textbf{Isotropic:}  
    This model augments the scalar viscosity and thermal conductivity using neural networks. The transport coefficients are defined as
    \begin{equation}
        \mu_I(\textbf{U};\theta) = \mu_0(1+f_{1I}(\textbf{U};\theta)), \quad \lambda_I(\textbf{U};\theta) = \lambda_0(1+f_{2I}(\textbf{U};\theta)),
        \label{eq: Scalar}
    \end{equation}
    where \( f_{1I} \) and \( f_{2I} \) are outputs of a shared neural network \( f_I(\textbf{U};\theta) \). This approach builds on prior work by Nair \emph{et al.}~\cite{nair2023deep} and assumes isotropy in all directions.

  \item \textbf{Anisotropic:}
    In continuum mechanics, the viscous stress tensor $\tau_{ij}$ for a Newtonian fluid can be expressed in terms of the symmetric strain-rate tensor $S_{kl}$ and a fourth-order viscosity tensor $\mu_{ijkl}$,
    \begin{equation*}
      \tau_{ij}=\mu_{ijkl}S_{kl},
    \end{equation*}
    which accounts for the fully anisotropic coupling between the viscous stress and strain rate.
    We reduce this to a diagonal anisotropic dependence by assuming negligible off-diagonal components of $\mu_{ijkl}$ (no coupling between shear and normal strain-rate components).
    We also reduce the fully anisotropic thermal conductivity tensor \cite{powers2004necessity} to a diagonal tensor by assuming negligible off-diagonal components.
    The diagonal viscosity and thermal conductivity tensors expressed in Voigt form~\cite{gurtin2010mechanics} are
    \begin{equation}
        {\mu}_\mathrm{A} = 
        \begin{bmatrix}
            \mu_{xx} & 0 & 0 \\
            0 & \mu_{yy} & 0 \\
            0 & 0 & \mu_{xy}
        \end{bmatrix}, \quad
        {\lambda}_\mathrm{A} = 
        \begin{bmatrix}
            \lambda_x & 0 \\
            0 & \lambda_y
        \end{bmatrix},
    \end{equation}
    where the neural network outputs $\mathbf{f}_{1,\mathrm{A}}$ and $\mathbf{f}_{2,\mathrm{A}}$ provide the anisotropy of the augmented transport coefficients:
    \begin{equation}
        {\mu}_\mathrm{A}(\textbf{U};\theta) = \mu_0(1+\mathbf{f}_{1,\mathrm{A}}(\textbf{U};\theta)), \quad {\lambda}_\mathrm{A}(\textbf{U};\theta) = \lambda_0(1+\mathbf{f}_{2,\mathrm{A}}(\textbf{U};\theta)).
        \label{eq: Aniso}
    \end{equation}
    The viscous stress tensor is obtained from the anisotropic viscosity tensor as
    \begin{equation}
    \mathbf{\tau}^{(v)} = 
    \begin{bmatrix}
      \tau_{11} \\
      \tau_{22} \\
      \tau_{12}
    \end{bmatrix} = {\mu}_\mathrm{A} 
    \left( 
    \begin{bmatrix}
        S_{11} \\
        S_{22} \\
        2S_{12}
    \end{bmatrix}
    - \frac{1}{3} (S_{11} + S_{22}) 
    \begin{bmatrix}
        1 \\
        1 \\
        0
    \end{bmatrix}
    \right),
    \label{eq:visc_stress_voigt}
    \end{equation}
    where $S^{(v)}=[S_{11},S_{22},2S_{12}]^\top$
    is the 2D strain-rate tensor in Voigt form. The subtraction of the isotropic component enforces the trace-free property of the stress tensor.
    The anisotropic heat-flux vector is
    \[
        \mathbf{q} = -{\lambda}_\mathrm{A} \, \nabla T.
    \]

    \item \textbf{Anisotropic-TF:}  
    To impose the physically motivated constraint that the viscous stress tensor for monatomic gases should be trace-free (TF), we modify $\mu_\mathrm{A}$ by setting the normal components equal, \( \mu_{xx} = \mu_{yy} \), such that the resulting viscous stress tensor \eqref{eq:visc_stress_voigt} is trace-free. The modified diagonal viscosity tensor is
    \begin{equation}
        {\mu}_{\mathrm{TF}} = 
        \begin{bmatrix}
            \mu_{xx} & 0 & 0 \\
            0 & \mu_{xx} & 0 \\
            0 & 0 & \mu_{xy}
        \end{bmatrix}.
    \end{equation}
    Similar to the preceding diagonal viscosity tensor $\mu_\mathrm{A}$, we obtain this trace-free-enforcing viscosity tensor from neural network outputs as ${\mu}_{\mathrm{TF}}(\textbf{U};\theta) = \mu_0(1+\mathbf{f}_{1,\mathrm{TF}}(\textbf{U};\theta))$, with the thermal conductivity unchanged from $\lambda_\mathrm{A}$. The viscous stress and heat flux are obtained as before.
\end{enumerate}

The inputs to each neural network, discussed in Section~\ref{sec:NN}, are selected to ensure Galilean and rotational invariance, which are essential for ensuring generalization across different flow configurations.
Modeling strategy (a) has been successfully applied to 1D viscous shocks in argon~\cite{nair2023deep,nair2023entropy}, and its performance is compared to that of (b) and (c) in Section~\ref{sec:Results}. The functional form of the neural networks and their training are described in Section~\ref{sec:NN}.

\subsubsection{Entropy Constraints} \label{sec:entropy}

Augmenting the viscous stress and heat flux in the momentum and energy equations using neural networks introduces the risk of violating fundamental physical laws. In particular, unconstrained model outputs may result in globally negative entropy production, which violates the second law of thermodynamics and can cause numerical instability or nonphysical predictions. To mitigate this, we incorporate thermodynamic constraints directly into the learning process. Specifically, we enforce the Clausius–Duhem inequality, which ensures nonnegative irreversible entropy generation~\cite{powers2016combustion}. The entropy production rate $\mathcal{\dot{I}}$ is expressed as
\begin{equation} \label{irrRevEnt} 
    \mathcal{\Dot{I}} = -\frac{1}{T^2} \bq(\bU;\theta) \cdot \nabla T + \frac{1}{T} {\tau}(\bU;\theta) : \nabla \bu^\top \ge 0, 
\end{equation}
where the {colon operator denotes the double contraction (Frobenius inner product) between two second-order tensors.}

A ``strong'' form of the entropy constraint would be to impose conditions on the model outputs such that each term in (\ref{irrRevEnt}) is nonnegative. A ``weak'' constraint for the entropy inequality would be to incorporate $\mathcal{\Dot{I}}$ into the training loss function \eqref{e:loss} to penalize  violations of the second law via the embedded optimization. Previous work in one-dimensional viscous shock problems~\cite{nair2023deep} found the strong constraint to yield better predictive performance and greater numerical stability. Accordingly, for the two-dimensional cases in this study, we adopt the strong constraint approach by enforcing positivity of the predicted transport coefficients during training, {which automatically satisfies the Clausius–Duhem inequality.}

\subsubsection{Neural Network Inputs and Architecture} \label{sec:NN}

The neural network receives input features that are designed to preserve key physical invariances, notably Galilean invariance and rotational invariance. At each mesh point, the local input vector is
\[
\hat z = \left[ |\overline{\nabla \rho}|, |\overline{\nabla p}|, |\overline{\nabla T}|, \text{Eigs}(\overline{S}), \overline{\rho}, \overline{p}, \overline{T} \right],
\]
where $\text{Eigs}(\overline{S})$ denotes the eigenvalues of the normalized strain-rate tensor, {which are computed by solving the characteristic equation of the tensor using \emph{PyTorch}'s \texttt{torch.linalg.eig} function, which returns the complex eigenvalues and eigenvectors of a general square matrix. For symmetric real-valued tensors such as the strain-rate tensor, the imaginary components of the eigenvalues are zero.} The neural network inputs are normalized as
\begin{equation}
  \overline{\phi} = \frac{\phi}{\phi_{\infty}}, \qquad | \overline{\nabla \phi} | = \frac{\Lambda_{HS} |\nabla \phi|}{\phi},
  \label{eq:norm}
\end{equation}
where $\phi_\infty$ is the respective freestream value and \(\Lambda_{HS}\) is the local hard-sphere mean free path {(\ref{Eq: Mean-free path})}. The gradient normalization \eqref{eq:norm} serves two roles: it normalizes gradients to yield dimensionless, Knudsen number-like input features that characterize local rarefaction, and it enhances the network’s sensitivity in regions where continuum assumptions break down, such as shock layers and near-wall boundary regions, by activating the model more prominently where nonequilibrium effects are significant.

\begin{figure}
    \centering \includegraphics[width=0.65\textwidth]{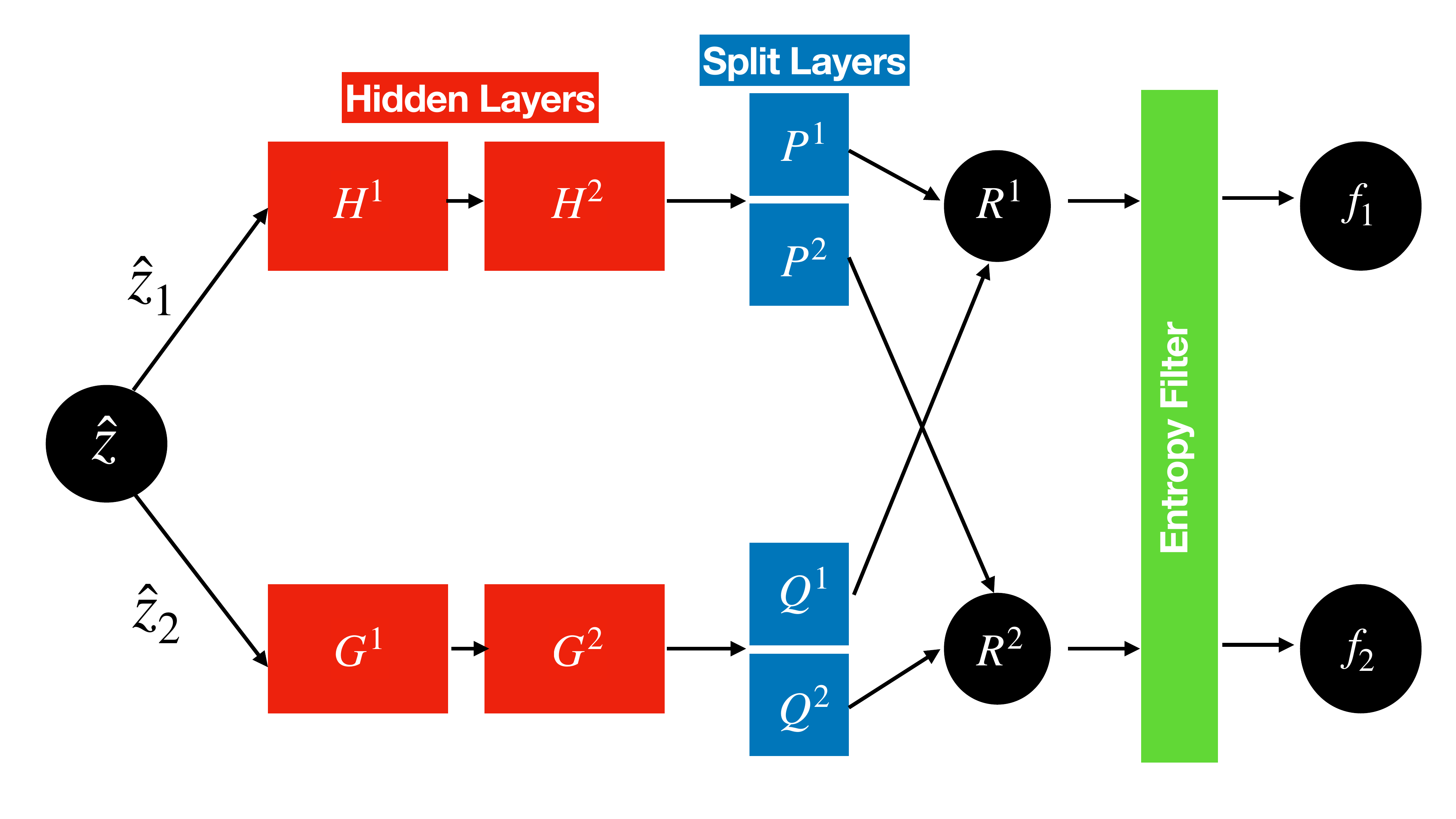} 
    \caption{Schematic of the neural network architecture used for augmented transport models.
    } 
    \label{fig:ArchiSchem} 
\end{figure}

 Figure~\ref{fig:ArchiSchem} illustrates the architecture of the neural network used for transport coefficient modeling, which is structured around a dual-branch, DeepONet-inspired architecture~\cite{lu2021learning}, where the input vector \(\hat{z}\) is split into two subspaces, one containing local gradients ($\hat{z}_1$) and the other containing the thermodynamic state variables ($\hat{z}_2$). The neural network architecture is
\begin{align}
    H^1 &= \sigma_1(W_1 \hat{z}_1 + b_1) &
    G^1 &= \sigma_1(W_2 \hat{z}_2 + b_2), \nonumber \\
    H^2 &= \sigma_1(W_3 H^1 + b_3) &
    G^2 &= \sigma_1(W_4 G^1 + b_4), \nonumber \\
    [P^1, P^2] &= W_5 H^2 + b_5 &
    [Q^1, Q^2] &= W_6 G^2 + b_6, \nonumber \\
    R^1 &= P^1 \cdot Q^1 & R^2 &= P^2 \cdot Q^2, \nonumber \\
    [{f}_1, {f}_2] &= \text{Entropy Filter}( W_7 [R^1, R^2] + b_7) + 1,
    \label{eq:NN_arch}
\end{align}
with outputs \([{f}_1, {f}_2]\). Depending on the given model, these are either local scalar-valued quantities (for the Isotropic model) or local vector-valued quantities (for the Anisotropic and Anisotropic-TF models).
The activation function \(\sigma_1(\cdot)\) is the exponential linear unit (ELU),
\begin{equation*}
\sigma_1(x) = \begin{cases}
\alpha(\exp(x) - 1), & x \leq 0, \\
x, & x > 0,
\end{cases}
\end{equation*}
with $\alpha=1$. In \eqref{eq:NN_arch}, the element-wise multiplications \(R_1 = P_1 \cdot Q_1\) and \(R_2 = P_2 \cdot Q_2\) represent learned nonlinear bases that capture interactions between local gradients and thermodynamic states while preserving separability between the two sets of outputs (viscosities and thermal conductivities). Each hidden layer $i$ has thirty weights $W_i$ and biases $b_i$, giving a total of $N_\theta=210$ learnable parameters
\[
\theta = \{ W_i, b_i \}_{i=1}^7
\]
over the seven hidden layers. The entropy filter is defined as
\begin{equation}
  \text{Entropy Filter($x$)} = \sigma_1(x) + 1
\end{equation}
to ensure positivity of the outputs.
The two separate neural network branches ensure that independent nonlinear {approximations} are learned for the modeled viscosities and thermal conductivities, thereby enabling the model to represent potentially distinct dependencies on the thermodynamic state variables their gradients, and the eigenvalues of the strain-rate tensor.

\subsection{Boltzmann Distribution Function Wall Models} \label{Sec : Boltzmann wall models}

As we show in Section~\ref{sec:Results}, the transport-coefficient models are successful in improving NSF predictions of transition-continuum boundary layer flows away from the wall but are not designed to improve predictions near the wall.
The first-order boundary conditions (Eqs.~\ref{eq: Slip-velocity} and \ref{eq: Temp-jump}) are derived using the Chapman--Enskog expansion of the Boltzmann equation by assuming small deviations from the Maxwell--Boltzmann velocity distribution function
\begin{equation}
    \mathbf{f}_0 (\mathbf{c}, \mathbf{u}) = \left( \frac{m}{2 \pi k_B T} \right)^\frac{3}{2} \text{exp} \left[ 
   \frac{m \left(\mathbf{c} - \mathbf{u} \right)^2}{2 k_B T} \right],
\end{equation}
where $\mathbf{c}$ is the particle velocity, $m$ is the molecular mass, and $k_B$ is the Boltzmann constant. Assuming a Maxwell--Boltzmann distribution can be problematic in regions of continuum breakdown, as the velocity distributions in these areas can be bimodal due to strong nonequilibrium effects. The assumption fails to capture the nature of the velocity distribution near walls or in shock layers, where significant deviations from equilibrium occur. A more capable closure is needed to provide accurate nonequilibrium boundary conditions.

{We model the near-wall nonequilibrium distribution function as a product of spatially independent distribution functions},
\begin{equation}
  \mathbf{f}(\mathbf{c}, \mathbf{x}) \approx \prod_{i=1}^{3} \tilde{\mathbf{f}}_i(c_i, \mathbf{x}),
\end{equation}
which assumes that the particle velocity components are uncorrelated. We model the directional distribution functions using a convex combination of skewed Gaussian distributions,
\begin{equation}
\tilde{\mathbf{f}}_i(c_i, \mathbf{x}) \approx \lambda_i \tilde{\mathbf{f}}_i^1(c_i, \mathbf{x}) + (1 - \lambda_i) \tilde{\mathbf{f}}_i^2(c_i, \mathbf{x}),
\end{equation}
where each skewed Gaussian distribution is given by
\begin{equation}
\tilde{\mathbf{f}}_i^k(c_i, \mathbf{x}) = \frac{2}{\sigma_i^k}\phi\left(\frac{c_i - \mu_i^k}{\sigma_i^k}\right)\Psi\left(\alpha_i^k\frac{c_i - \mu_i^k}{\sigma_i^k}\right),
\end{equation}
in which the standard Gaussian distribution  and cumulative distribution are
\begin{equation}
\phi(z) = \frac{1}{\sqrt{2\pi}}e^{-\frac{z^2}{2}}, \quad \Psi(z) = \int_{-\infty}^{z}\phi(t)dt.
\end{equation}
Each modeled distribution function therefore involves 21 free parameters
\begin{equation}
  \Theta_i^w(\bU) = [\sigma_i^k, \mu_i^k, \alpha_i^k, \lambda_i],
\end{equation}
which we permit to depend on the flow solution $\bU$ so that the bimodality and skewness of the modeled distribution function can vary across the domain. We obtain $\Theta_i^w$ using an additional neural network with trainable parameters $\theta_w$,
\begin{equation}
  \Theta_i^w(\bU;\theta_w) = \mathbf{f}_w(\hat{z};\theta_w),
\end{equation}
the structure of which is given in Sec.~\ref{sec:wall_NN} and is separate from that of the transport-coefficient neural networks, but which the same local inputs $\hat{z}$.

Analytical expressions for the first and second moments of each skewed Gaussian distribution are obtained as
\begin{align}
\mathbf{I1}_i^k &= \int_{-\infty}^{\infty}(c_i - u_i)\tilde{\mathbf{f}}_i^kdc_i = \mu_i^k + \sigma_i^k\delta_i^k\sqrt{\frac{2}{\pi}} - u_i,\\
\mathbf{I2}_i^k &= \int_{-\infty}^{\infty}(c_i - u_i)^2\tilde{\mathbf{f}}_i^kdc_i = (\sigma_i^k)^2\left(1 - \frac{2(\delta_i^k)^2}{\pi}\right) + \left(\mu_i^k + \sigma_i^k\delta_i^k\sqrt{\frac{2}{\pi}} - u_i\right)^2.
\end{align}
The moments of the directional distribution functions are obtained using the convex combination
\begin{equation}
\mathbf{I1}_i = \mathbf{I1}_i^1 + \lambda_i\mathbf{I1}_i^2, \qquad \mathbf{I2}_i = \mathbf{I2}_i^1 + \lambda_i\mathbf{I2}_i^2,
\end{equation}
from which the
streamwise velocity is obtained as
\begin{equation}
  u(\mathbf{x}) = \iiint_{-\infty}^{\infty}\tilde{\mathbf{f}}_1(c_i, \mathbf{x})d\mathbf{c} = \mathbf{I1}_1,
  \label{eq:dist_u}
\end{equation}
and the temperature is 
\begin{equation}
  T(\mathbf{x}) = \frac{m}{3k}\iiint_{-\infty}^{\infty}c_i^2\tilde{\mathbf{f}}_i(c_i, \mathbf{x})d\mathbf{c} = \frac{m}{3k}\mathbf{I2}_1\mathbf{I2}_2\mathbf{I2}_3.
  \label{eq:dist_T}
\end{equation}
We use \eqref{eq:dist_u} and \eqref{eq:dist_T} in the present study to obtain the wall slip velocity $V_s$ and temperature $T_w$, replacing the standard velocity-slip/temperature-jump boundary conditions described in Sec.~\ref{sec:slipwall}.
This approach based on modeled distribution functions could be extended to direct calculation of heat fluxes and viscous stresses throughout the domain. However, a substantial number of additional constraints would be needed to ensure that the standard continuum thermodynamic principles would remain satisfied, as the transport models introduced in Sec.~\ref{sec:transport_model} ensure.

\subsubsection{Model Architecture and Inputs} \label{sec:wall_NN}

The inputs to the wall-model neural network are the same as those used for the transport coefficient model, \(\hat{z}\), defined in Section~\ref{sec:NN}. 
The architecture of the wall-model neural network is similar to that used in previous DPM studies \cite{sirignano2020dpm,macart2021embedded},
\begin{equation}
\begin{split}
    H^1 &= \sigma_1(W_1 \hat{z} + b_1), \\
    H^2 &= \sigma_1(W_2 H^1 + b_2), \\
    H^3 &= \sigma_1(W_3 H^2 + b_3), \\
    G   &= \sigma_2(W_G \hat{z} + b_G), \\
    H^4 &= G \odot H^3, \\
    \Theta^w_1 &= W_4 H^4 + b_4, \\
    \Theta^w_2 &= W_5 H^4 + b_5, \\
    \Theta^w_3 &= W_6 H^4 + b_6, \\
    [\Theta^w_1, \Theta^w_2, \Theta^w_3] &= \text{concat}(\Theta^w_1, \Theta^w_2, \Theta^w_3),
\end{split}
\end{equation}
where each \(\Theta^w_i\) denotes the set of output parameters associated with the modeled particle velocity distribution in the \(i^\text{th}\) spatial direction. We use ELU activation functions \(\sigma_1(\cdot)\) as before, and \(\sigma_2(\cdot)\) is the sigmoid function. The operator \(\text{concat}(\cdot)\) denotes the concatenation of the three output vectors along the feature dimension to produce the parameters of the modeled distribution functions.

\subsection{Neural Network Optimization}

The neural network closures introduced in Sections~\ref{sec:transport_model} and \ref{Sec : Boltzmann wall models} are embedded within the governing PDE system and are trained to minimize the discrepancy between model predictions and DSMC target data.
The DSMC target data is generated using the \emph{SPARTA} solver \cite{plimpton2019direct}.
We now discuss the optimization (training) of the neural network transport and wall models.

Both sets of models use an integrated loss (objective) function comprising the normalized mean-squared error of the density, velocity, and temperature,
\begin{equation}
J(\theta)=\frac{1}{2} \int_{\Omega}\left[\frac{1}{\rho_{\infty}^2}(\rho(\bx; \theta)-\rho_T)^2 + \frac{1}{u_{\infty}^2}\sum_{i=1}^{d}(u_i(\bx; \theta)-(u_T)_i)^2 + \frac{1}{T_{\infty}^2}(T(\bx; \theta)-T_T)^2 \right] d\bx,
\label{e:loss}
\end{equation}
where the domain of integration $\Omega$ spans the entire computational domain for transport models and is limited to the wall for wall models.
In \eqref{e:loss}, we denote the implicit dependence of the flow variables on the neural network parameters, and $\rho_T$, $u_T$, and $T_T$ are spatially local targets obtained from DSMC solutions. To emphasize learning in regions with strong nonequilibrium effects, the loss is computed in a clipped domain extending from $x = -0.1$ to $x = 1.0$ (with $x = 0$ corresponding to the leading edge of the wall) and from $y = 0$ to $y = 0.3$, thereby excluding much of the far-field equilibrium region. This loss-function formulation can be adapted to use conserved variables or derived quantities (e.g., pressure, energy, or wall fluxes) depending on the specific modeling goal. We next discuss the adjoint-based optimization method used to train the embedded neural networks in conjunction with the flow PDEs.

\subsubsection{Adjoint-Based Gradient Computation and Parameter Update}

Since the neural networks are embedded in the governing equations, computing gradients of the loss function with respect to the model parameters requires accounting for the implicit dependence of the flow solution on the neural network outputs. This is handled using an adjoint-based optimization approach \cite{sirignano2020dpm}.

Let $\bU = [\rho, \bu, T]^\top$ be the primitive state and {$\mathbf{F}_x=\nabla\cdot(\bF_d-\bF_c)$ the total flux divergence} as discretized by the PDE solver. The sensitivity of the loss function to the network parameters is
\begin{equation} 
    \nabla_{\theta} J = {\frac{\partial \mathbf{F}_x}{\partial \theta}}^\top \hat{\bU},
    \label{eq:loss_gradient}
\end{equation}
where $\hat{\bU} = [\hat{\rho}, \hat{\bu}, \hat{T}]^\top$ is the adjoint state, which is obtained by solving the steady-state adjoint equation
\begin{equation}
    \hat{\bU}^\top \frac{\partial \mathbf{F}_x}{\partial \bU} + \frac{\partial J}{\partial \bU} = 0.
    \label{eq:adjeq}
\end{equation}
We solve the adjoint system using the biconjugate gradient method~\cite{saad2003iterative}, which is efficient for the large, sparse Jacobian matrix of the discretized governing equations.

For consistency, the transport models and wall models (when used) are trained using the same optimization procedure. Each model (or combination of models) is trained for 60 optimization iterations, starting from a converged Navier--Stokes solution using the standard power-law viscosity and thermal conductivity \eqref{eq:sutherland} and the Maxwellian velocity-slip/temperature-jump boundary conditions (Eqs.~\ref{eq: Slip-velocity} and \ref{eq: Temp-jump}, i.e., without any ML augmentation). At each optimization step, the forward solver residuals are converged to \(10^{-8}\) to ensure accurate and consistent gradient evaluation, then the adjoint equation \eqref{eq:adjeq} is solved.

Once the adjoint variables are computed, the gradient $\nabla_\theta J$ is obtained using \eqref{eq:loss_gradient}, and the neural network weights are updated using  using the Adam optimizer~\cite{kingma2014adam}, a first-order gradient-based method with adaptive learning rates,
\begin{equation}
    \theta^{n+1} = \theta^{n} - \alpha^{n} \nabla_\theta J,
    \label{Eq: grad descent}
\end{equation}
where $\alpha^{n}$ is the learning rate at iteration $n$. 
The initial learning rate is \(\alpha^{0} = 10^{-4}\) and is adjusted adaptively using a learning rate decay schedule that promotes stable convergence \cite{nair2023deep}. The learning rate is halved whenever the loss falls below 90\% of a threshold according to the update rule
\begin{equation}    
    \alpha^{n+1} =
    \begin{cases}
    \gamma \alpha^{n} & \text{if } J^n \leq 0.9 J^n_{\text{threshold}},\quad J^{n+1}_{\text{threshold}} \gets 0.9 J^n_{\text{threshold}}, \\
    \alpha^{n} & \text{otherwise},
    \end{cases}
    \end{equation}
where \(\gamma = 0.5\) is a decay factor and \(J^0_{\text{threshold}} = J^0\) is the loss of the converged, non-ML-augmented Navier--Stokes solution.
{Algorithm \ref{alg:Training} summarizes the training process}.

\begin{figure}
    \centering
    \includegraphics[width=0.5\textwidth]{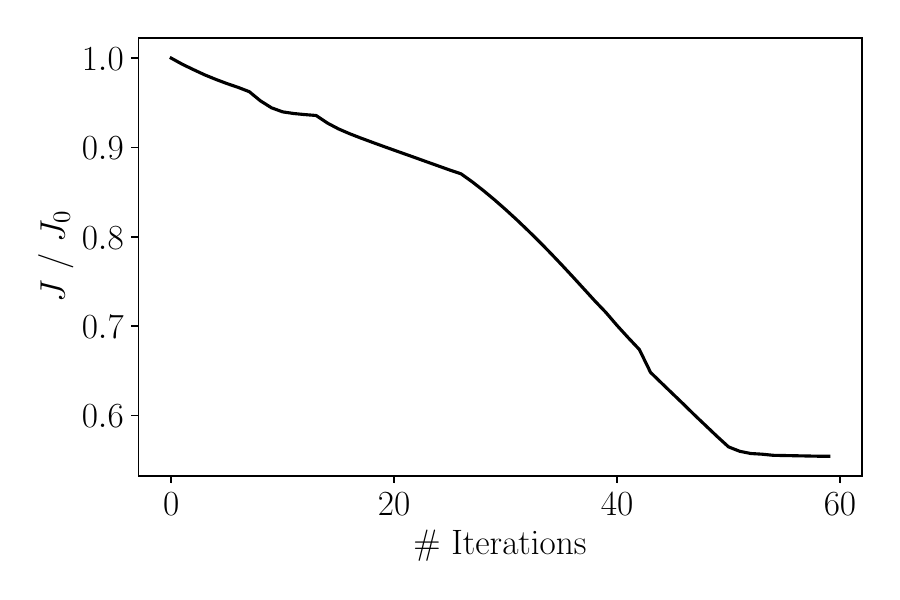} 
    \caption{Loss convergence of a co-optimized Anisotropic-TF transport model and distribution function-based wall model  for a \(M_\infty = 7\) flat-plate boundary layer.
    }
    \label{fig: LossConvergence}
\end{figure}

Figure~\ref{fig: LossConvergence} shows the convergence of the normalized loss, \(J / J^0\), as a function of optimization iterations for a co-optimized Anisotropic-TF model and neural network wall model. The steadily decreasing loss reaches approximately 50\% of its initial value after around fifty iterations, beyond which the loss plateaus, indicating that the model has approached a local minimum in the optimization landscape.

\begin{algorithm}
\caption{Training procedure for transport models and coupled transport--wall models}\label{alg:Training}
\begin{algorithmic}[1]

\State Restart $\mathbf{U}$ from unmodified Navier--Stokes solution
\State Randomly initialize neural network parameters $\theta^0$
\State Calculate loss of unmodified solution $J_0$
\State Set $\epsilon_\mathrm{rel}^0 = 1.0$
\For{$n \in [1, 60]$}
    \State Converge solution to steady-state 
    \State Solve adjoint equation \eqref{eq:adjeq} for $\hat{\bU}$
    \State Compute  $\nabla_{\theta} J(\bU(\theta^n))$ using \eqref{eq:loss_gradient}
    \State Update $\theta^{n+1}\gets\theta^n$ using gradient descent \eqref{Eq: grad descent}
    \State Calculate $\epsilon_\mathrm{rel}^{n+1}$
    \State Compute $\Delta \epsilon_\mathrm{rel}^{n+1} = \epsilon_\mathrm{rel}^{n+1} - \epsilon_\mathrm{rel}^{n}$
\EndFor
\end{algorithmic}
\end{algorithm}

\subsubsection{Training of Wall Models}

When used, the distribution function-based wall models are co-optimized alongside the transport models using the adjoint-based training framework and loss function described previously. Although the architecture of the wall-model neural network is relatively simple, a modified training strategy is employed to ensure stable convergence during co-optimization.

In early training stages, the wall model typically inaccurately predicts the wall velocity and temperature. These poor predictions more readily destabilize training simulations than inaccurate estimates of the transport properties, which easily leads to solver divergence and optimization failure. We therefore introduce a training stabilization strategy to mitigate this instability.
At optimization iteration $n$, let the imposed wall slip velocity in the training simulation be a weighted combination of the target DSMC bulk velocity at the wall and the neural network-modeled wall velocity,
\begin{equation}
    V_s^n = (1 - \lambda^n) V_{s,\text{DSMC}} + \lambda V_s(\bU;\theta_w^n),
\end{equation}
where \(\lambda^n\in[0,1]\) is a weighting factor, initially $\lambda^0=0$, with $\lambda^{n+1}=\min(\lambda^n+0.25,1)$ for $\bmod(n+1,10)=0$, i.e., every 10 iterations. An analogous weighting is applied to the wall temperature. The fact that $\lambda^n=1$ for $n\rightarrow\infty$ ensures that the DSMC target data is not directly used in the training simulation in later stages.

\begin{figure}
    \centering
    \includegraphics[width=0.49\textwidth]{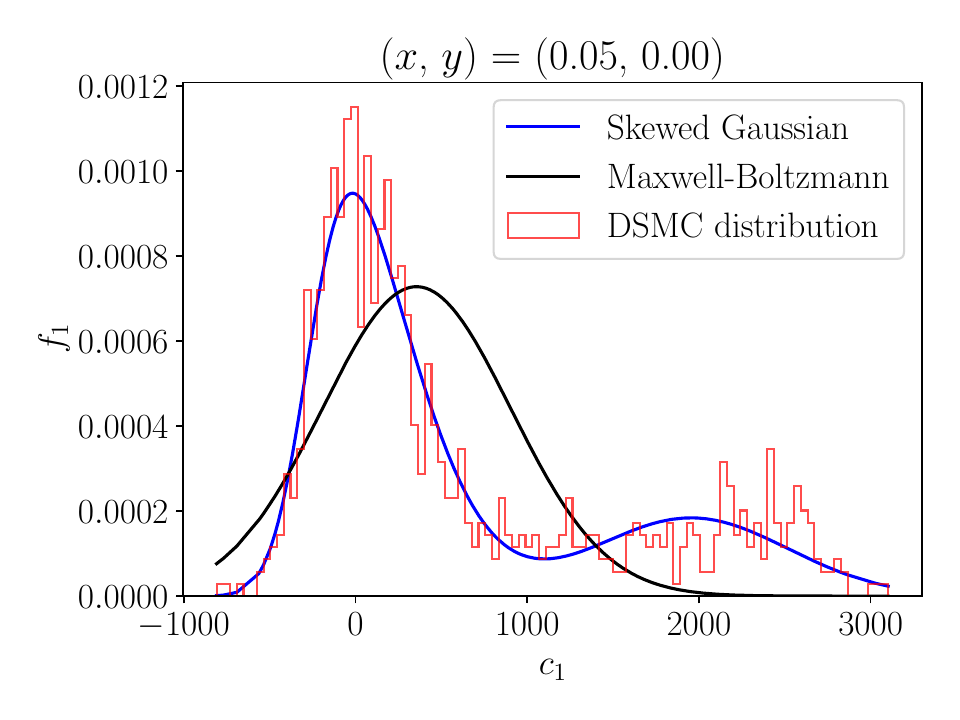} 
    \includegraphics[width=0.49\textwidth]{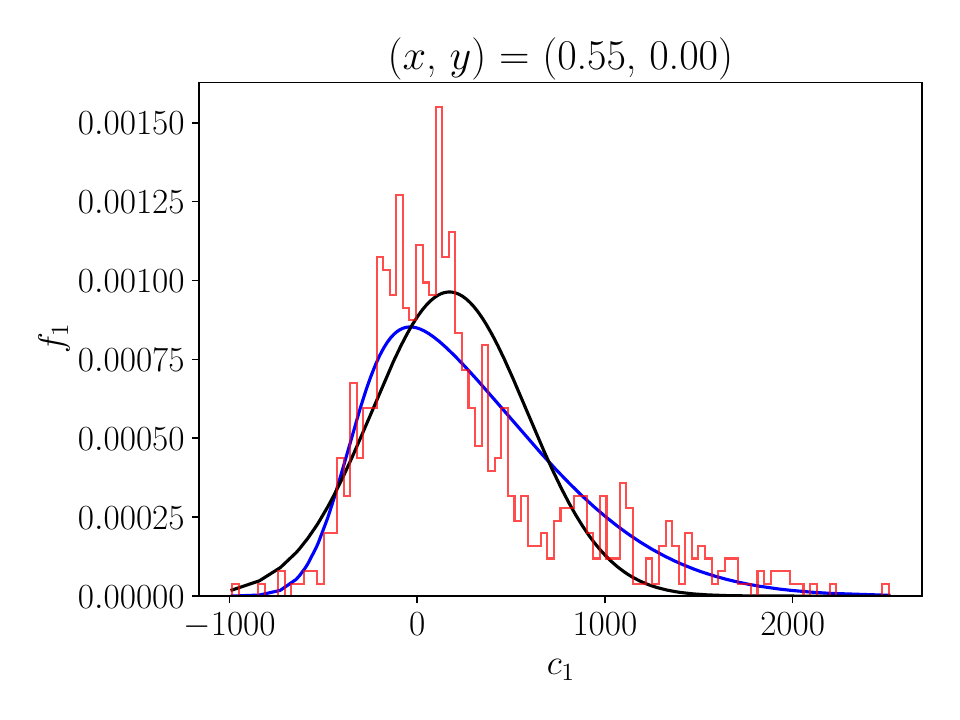} 
    \caption{Streamwise velocity distribution functions obtained from DSMC, the Maxwell--Boltzmann distribution function, and the modeled skewed-Gaussian distribution function at wall locations \(x = 0.05\)~m and \(x = 0.55\)~m for a \(M_\infty = 7\) flat-plate boundary layer.
    }
    \label{fig: Dist_compare}
\end{figure}

Figure~\ref{fig: Dist_compare} compares the DSMC-computed streamwise distribution function, the Maxwell--Boltzmann distribution function, and the learned skewed-Gaussian approximation at the wall. {The learned distribution function is trained to minimize the error in the continuum variables at the wall \eqref{e:loss} using the adjoint procedure, which ensures consistency with the continuum PDE system \eqref{eq:1D N-s}, and is not trained to directly minimize the mismatch with the DSMC-computed distribution function.}
{The scatter in the DSMC data arises from statistical noise inherent to particle-based methods, particularly in low-density, near-wall regions where sampling is challenging. This is a well-known limitation of DSMC and does not indicate issues with physical accuracy.} {We compare against the Maxwell--Boltzmann distribution as a canonical equilibrium reference. While the Chapman--Enskog distribution would provide a first-order nonequilibrium correction, it assumes near-continuum conditions and fails to represent the bimodal features seen in transitional regimes, making it less suitable as a baseline for the rarefied flow conditions considered here.} 

At streamwise position \(x = 0.05\) m, the DSMC distribution exhibits a distinctly bimodal structure, characteristic of strong nonequilibrium effects near the flat-plate leading edge, which the Maxwell--Boltzmann distribution fails to capture.
In contrast, the skewed Gaussian approximation captures both the bimodality and the peak locations. Though the model slightly underpredicts the secondary peak at \(x = 0.05\) m, it nonetheless demonstrates a strong capacity to approximate the nonequilibrium particle velocity distributions observed in rarefied boundary layers. This capability is crucial for accurately reconstructing wall behavior in transition-continuum regimes where classical assumptions break down.

\subsection{Parallel Training Across Multiple Knudsen and Mach Numbers} \label{sec:parallel_training}

To improve generalization across varying degrees of flow rarefaction, we implement a parallel training strategy in which the neural networks are trained simultaneously for multiple Knudsen numbers \(\mathrm{Kn} = \{0.3,0.6,1.2\}\) and Mach numbers  \(M_\infty = \{7, 12\}\).
{We use a domain-maximum Knudsen number $\max_\bx \mathrm{Kn}(\bx)$, where $\mathrm{Kn}(\bx) = \Lambda_{HS}(\bx)/L_c(\bx)$ is the local Knudsen number using the local mean free path  $\Lambda_{HS}$ and density gradient-based characteristic length scale $L_c$.}

Each parallel training case is simulated by a dedicated message-passing interface (MPI) process, which performs independent forward and adjoint solves during each optimization iteration. {All model copies are initialized identically and are synchronized across the MPI processes between optimization iterations to ensure consistency in parameter updates.}

Before optimizing, the per-process parameter gradients \(\nabla_\theta J_i\) are aggregated using a weighted averaging scheme~\cite{kryger2024optimizationsecondordertransportmodels},
\begin{equation}    
\nabla_\theta J = \sum_{i=1}^{N} \frac{\chi_i}{\sum_{j=1}^{N} \chi_j} \nabla_\theta J_i,
\end{equation}
where \(\chi_i\) denotes the weight assigned to the \(i^\text{th}\) MPI process. In this work, we set \(\chi_i = 3\), 5, and 10 for \(\mathrm{Kn} = 0.3\), \(0.6\), and \(1.2\), respectively. This weighting emphasizes losses for higher-Knudsen number cases, for which deviations from continuum behavior are more pronounced and ML-augmented closure models are more critical. The weights $\chi_i$ are unchanged for different Mach numbers.

This strategy enables models trained for a range of conditions to learn a unified set of parameters, sensitive to physical trends across rarefaction levels, while also controlled for the relative influence of each constituent simulation in the optimization process. The parallelization using MPI ensures scalability, while the weighted gradient aggregation enhances robustness across nonequilibrium flow conditions.

\section{Numerical Tests} \label{sec:Results}

{We now assess the performance of the modeling framework described in Sections~\ref{sec:Governing Equations} and~\ref{sec:ML Closure} for two-dimensional hypersonic flat-plate boundary layer flows. The two-dimensional flat-plate configuration serves as a critical test case for assessing the generalization capability of the proposed models in higher-dimensional, nonuniform, and practically relevant hypersonic flow regimes than the one-dimensional shocks reported in previous literature \cite{nair2023deep}.}

{Initially, we train ML closure models using data} for freestream Mach number $M_\infty = 7$, with freestream velocity $U_{\infty} = 2457.16$ m/s, temperature $T_{\infty} = 300$ K, and density $\rho_{\infty} = 1.69 \times 10^{-5}$ kg/m$^3$. These conditions represent moderate rarefaction, for which the Navier--Stokes-predicted shock and boundary layer deviate significantly from the DSMC solution, thus providing a suitable environment in which to evaluate nonequilibrium transport corrections.

To evaluate the extrapolation capability of the trained models, we conduct out-of-sample tests for two additional freestream Mach numbers, $M_\infty = 3$ and $M_\infty = 10$. The corresponding freestream velocities are $U_{\infty} = 1230.41$ m/s and $U_{\infty} = 3500.15$ m/s, respectively, with the freestream temperature and density fixed to isolate the influence of velocity variation. The range of flow conditions spanned by these out-of-sample tests allows us to assess the robustness and physical consistency of the learned closures.

\begin{figure}
    \centering
    \includegraphics[width=0.67\textwidth]{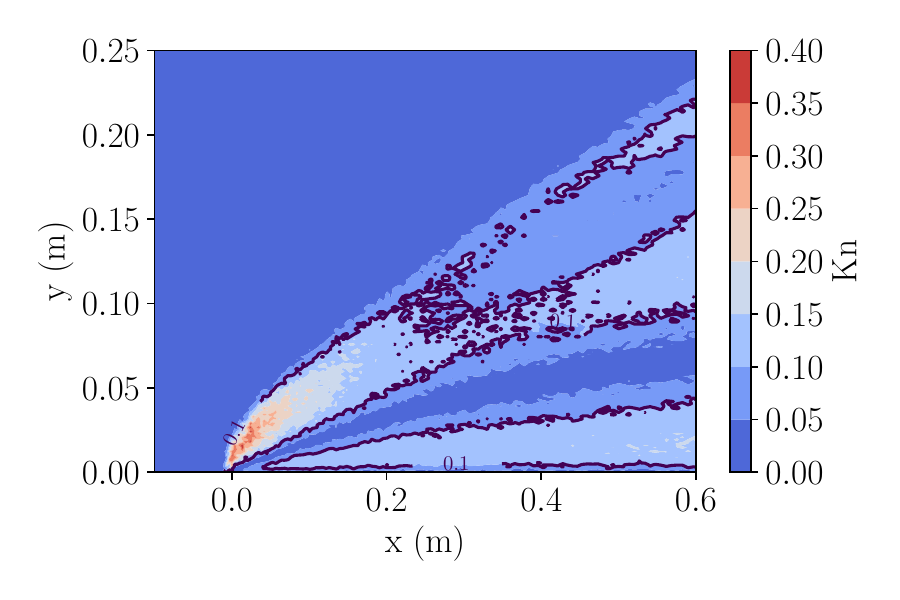} 
    \caption{Local density gradient-based Knudsen number for freestream conditions $U_{\infty} = 2457.16$ m/s, $T_{\infty}=300$ K and $\rho_{\infty} = 9.89013^{-5}$ kg/$m^3$, showing continuum breakdown ($\mathrm{Kn}=0.1$ isocontours) behind the shock and in the boundary layer. 
    }
    \label{fig: LocallKn}
\end{figure}

Figure~\ref{fig: LocallKn} shows the local Knudsen number distribution for the in-sample $M_\infty = 7$ case. Continuum breakdown, typically identified by $\mathrm{Kn} > 0.1$, is clearly visible in regions of strong compression within the shock and boundary layer. These localized nonequilibrium zones motivate the need for accurate, spatially aware closure models capable of adapting to varying degrees of rarefaction throughout the domain.

\subsection{Augmented Transport Models} \label{Sec : Augmented viscosity models}

We first assess the predictive accuracy and generalizability of the three augmented transport models introduced in Section~\ref{sec:ML Closure}(the Isotropic, Anisotropic, and Anisotropic–TF models) using the standard Maxwell boundary conditions, i.e., without using distribution function-based wall models.
Comparisons are made to the DSMC-evaluated macroscopic density, streamwise velocity, temperature, viscous stress, and heat flux, {for domain-maximum Knudsen number $\mathrm{Kn}=0.3$}. The results are used to evaluate the effectiveness of each model in capturing continuum breakdown effects and improving over the unmodified Navier--Stokes predictions.

\subsubsection{Comparison of Errors}

The relative error metric used to assess the accuracy of the different models is defined as
\begin{equation}
    \epsilon_{r,\phi} = \frac{|\phi|_2^N}{(|\phi|_2^N)_0},
\end{equation}
where $|\phi|_2^N$ denotes the normalized domain \(L_2\) error for a primitive variable $\phi$, 
\begin{equation}
    |\phi|_2^N = \left( \frac{\phi}{\phi_T} - 1 \right)^2,
    \label{eq: norm_L2}
\end{equation}
where $\phi_T$ is the corresponding DSMC target, and $(|\phi|_2^N)_0$ is the corresponding error obtained from the unmodified Navier--Stokes solution.

\begin{figure}
    \centering
    \includegraphics[width=0.9\textwidth]{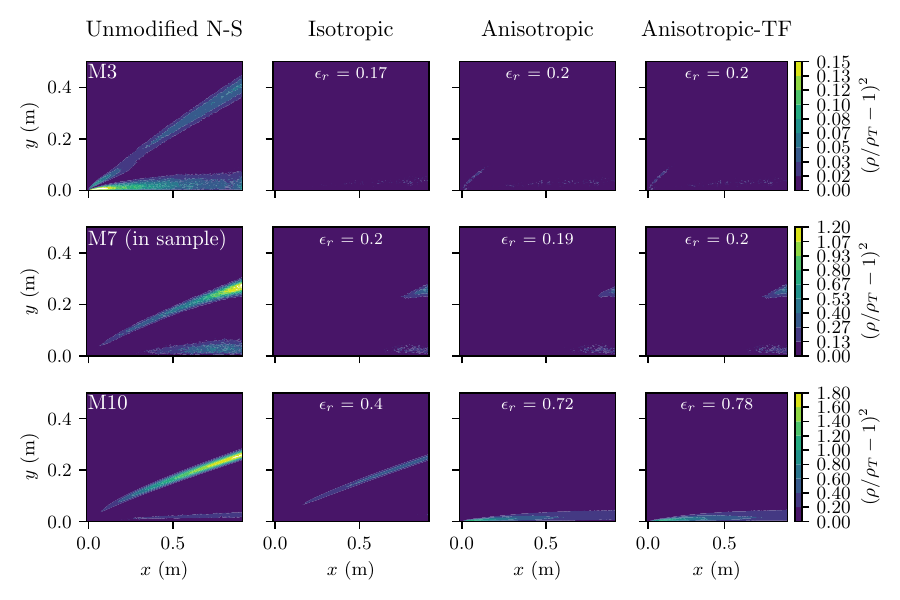} 
    \caption{Relative density errors  with respect to DSMC targets. $\epsilon_r$ is the domain L2 relative error.
    }
    \label{fig: LD_Dens_Compare}
\end{figure}

\begin{figure}
    \centering
    \includegraphics[width=0.9\textwidth]{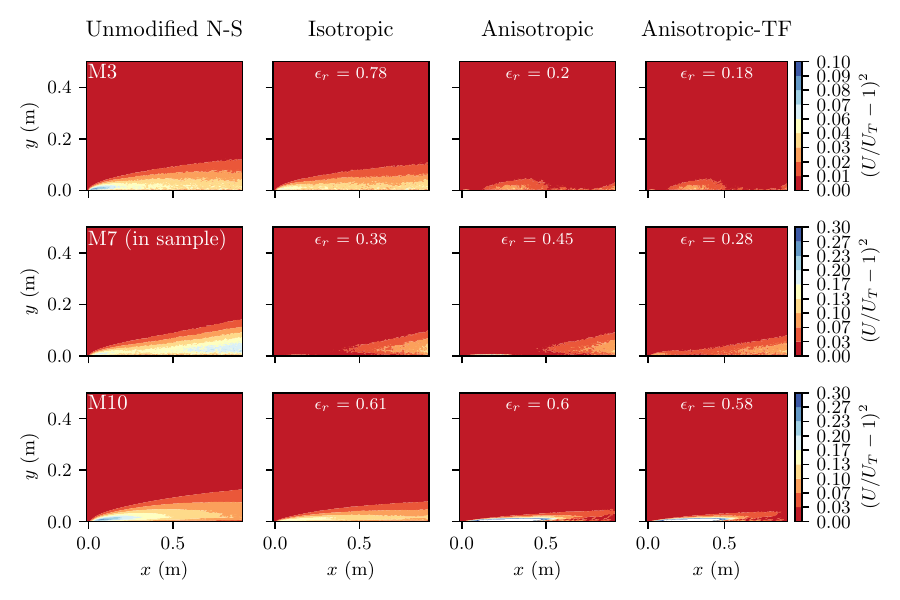} 
    \caption{Relative streamwise velocity errors with respect to DSMC targets. $\epsilon_r$ is the domain L2 relative error.}
    \label{fig: LD_UVel_Compare}
\end{figure}

\begin{figure}
    \centering
    \includegraphics[width=0.9\textwidth]{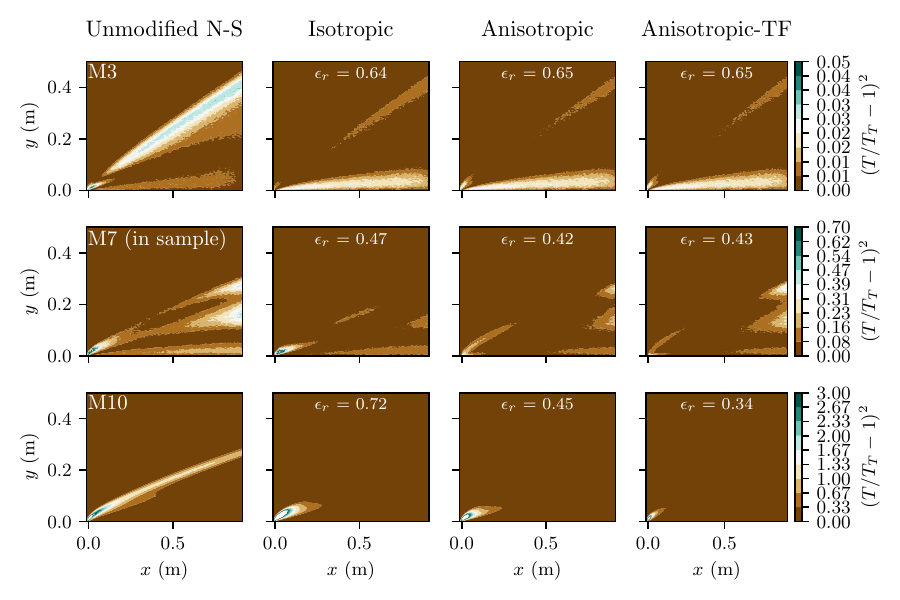} 
    \caption{Relative temperature errors with respect to DSMC targets. $\epsilon_r$ is the domain L2 relative error.}
    \label{fig: LD_Temp_Compare}
\end{figure}

Figures~\ref{fig: LD_Dens_Compare}, \ref{fig: LD_UVel_Compare} and \ref{fig: LD_Temp_Compare} compare the relative error of the predicted density, streamwise velocity, and temperature fields using the unmodified transport model to the predicted fields using $M_\infty=7$, $\mathrm{Kn}=0.3$-optimized Isotropic, Anisotropic, and Anisotropic-TF models. All three models reduce the errors across all fields compared to the no-model solution for both the in- and out-of-sample freestream Mach numbers.

Each of the optimized models improves predictive accuracy over the unmodified Navier--Stokes equations. While the isotropic model typically produces the lowest error in the predicted density field (Figure~\ref{fig: LD_Dens_Compare}),
the Anisotropic and Anisotropic-TF models exhibit significantly lower streamwise velocity error (Figure~\ref{fig: LD_UVel_Compare}).
For temperature (Figure~\ref{fig: LD_Temp_Compare}), the anisotropic models slightly outperform the isotropic model, particularly at the out-of-sample $M_\infty=10$.

\begin{figure}
    \centering
    \includegraphics[width=0.49\textwidth]{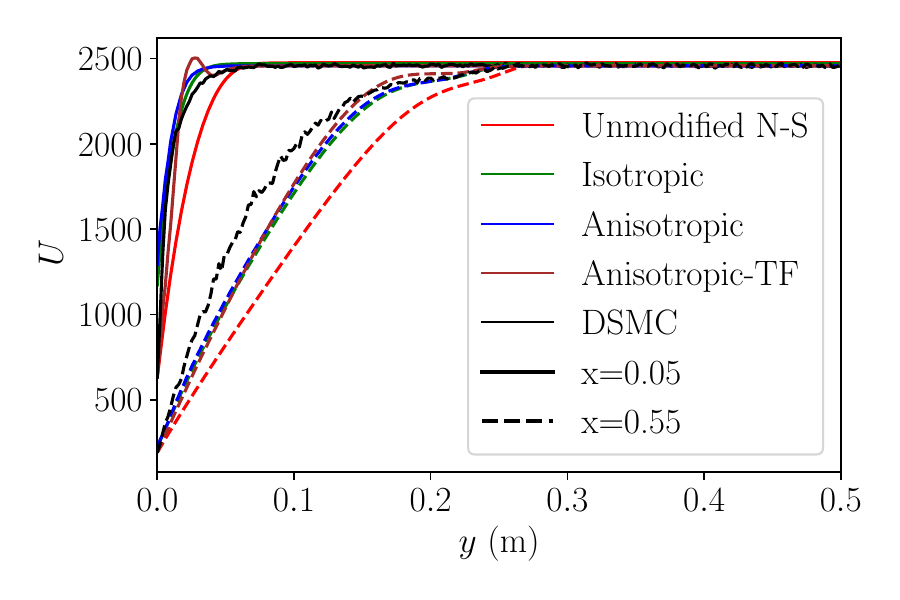} 
    \includegraphics[width=0.49\textwidth]{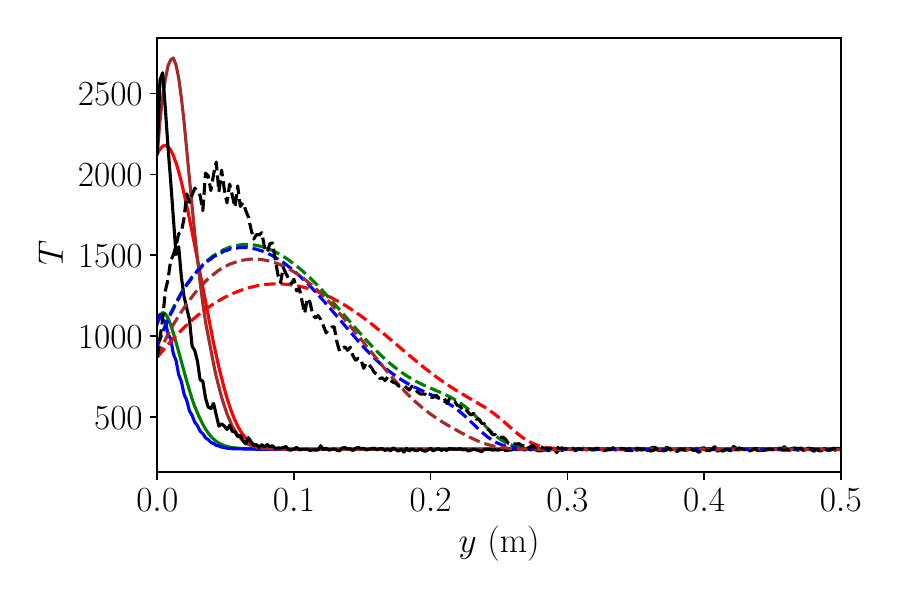} 
    \caption{Streamwise velocity and temperature predictions in the wall-normal direction at two points on the wall at ($x=0.05$ m and $x=0.55$ m) for the in-sample $M_\infty=7$, $\mathrm{Kn}=0.3$  case.}
    \label{fig: LD_U_T_prof_compare}
\end{figure}

Figure~\ref{fig: LD_U_T_prof_compare} shows wall-normal streamwise velocity and temperature profiles at streamwise locations \(x = 0.05\) m and \(x = 0.55\) m.
Near the leading edge, at \(x = 0.0.5\) m, all three ML-augmented predictions closely match the DSMC-evaluated velocity profile, improving upon the unmodified Navier--Stokes predictions,  and this improvement continues at the downstream location \(x = 0.55\) m.
However, the temperature profiles exhibit notable differences between the models. At \(x = 0.05\) m, the unmodified Navier--Stokes solution and the Isotropic model-augmented solution significantly underpredict the peak temperature, while the anisotropic models more closely match the peak temperature, though their wall temperature gradients do not match those of the DSMC profile, likely due to the first-order Maxwell boundary conditions. At \(x = 0.55\) m, all models underpredict the peak temperature and the wall temperature gradient, though the Anisotropic model-augmented solutions more closely match the DSMC profile. Overall, the ML-augmented solutions are most accurate near the leading edge and less so downstream.

\subsubsection{Viscous Stress and Heat Flux}

\begin{figure}
    \centering
    \includegraphics[width=0.49\textwidth]{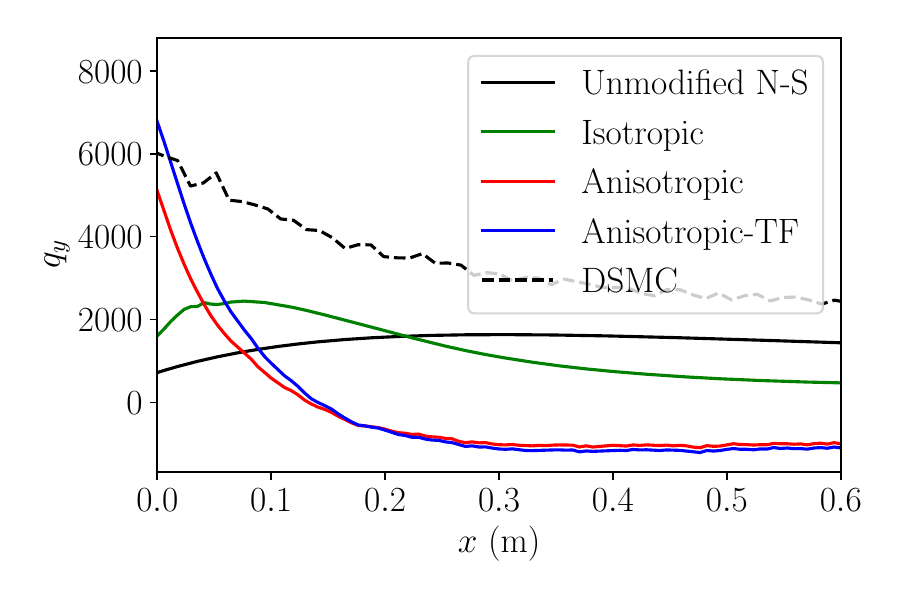} 
    \includegraphics[width=0.49\textwidth]{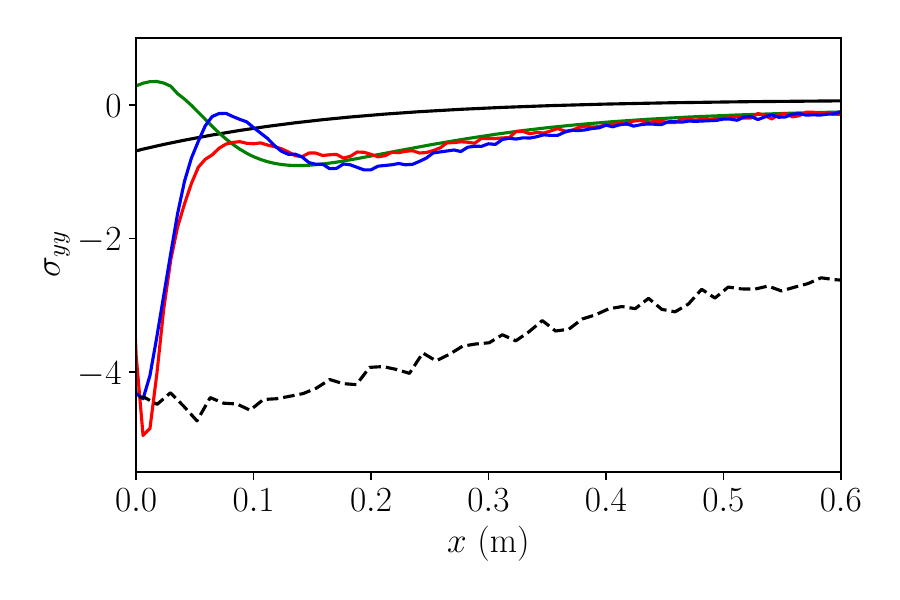} 
    \caption{Wall-normal heat flux and viscous stress for the in-sample $M_\infty=7$, $\mathrm{Kn}=0.3$ case.
    } 
    \label{fig: LD_sigma_q_Compare}
\end{figure}

Figure~\ref{fig: LD_sigma_q_Compare} shows the unmodified and ML-augmented Navier--Stokes predictions of the wall-normal heat flux and viscous stress, compared to the DSMC-evaluated profiles, along the wall. The unmodified Navier--Stokes solution significantly underpredicts the peak heat flux and fails to capture the overall trends, and the Isotropic model slightly improves upon this. The Anisotropic and Anisotropic-TF models significantly improve the predicted heat flux at the leading edge (i.e., before the boundary layer develops), but do not improve predictive accuracy of these important wall quantities downstream. This demonstrates the significant influence of the slip-wall boundary condition, which here remains unchanged. These deficiencies are addressed next using the modeled distribution function-based wall model.

\subsection{Coupled Transport and Boundary Condition Models}

We now evaluate the impact of incorporating the Boltzmann distribution function-based wall model into the Anisotropic-TF transport model, which overall offers the best performance as shown in the previous section. The comparison is performed using the same error metrics employed previously.

\subsubsection{Comparison of Errors}

\begin{figure}
    \centering
    \includegraphics[width=0.9\textwidth]{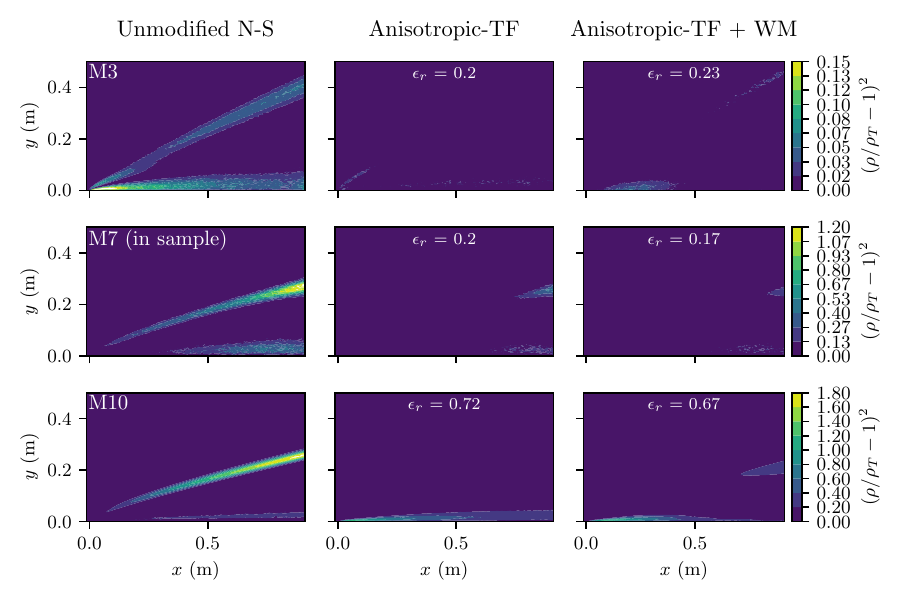} 
    \caption{Relative density errors  with and without the ML wall model (WM) with respect to DSMC targets.}
    \label{fig: BW_Dens_Compare}
\end{figure}

\begin{figure}
    \centering
    \includegraphics[width=0.9\textwidth]{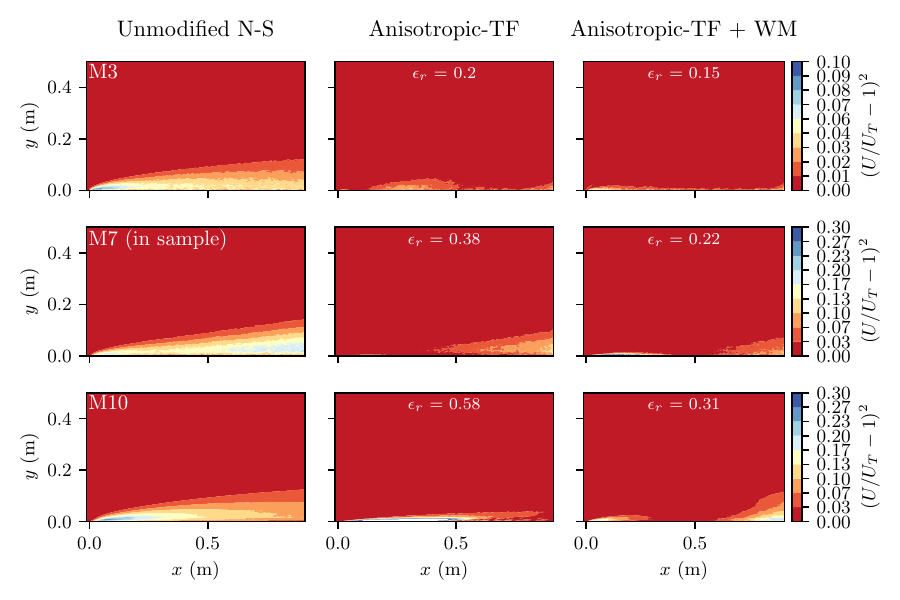} 
    \caption{Relative streamwise velocity errors  with and without the ML wall model (WM)  with respect to DSMC targets.}
    \label{fig: BW_UVel_Compare}
\end{figure}

\begin{figure}
    \centering
    \includegraphics[width=0.9\textwidth]{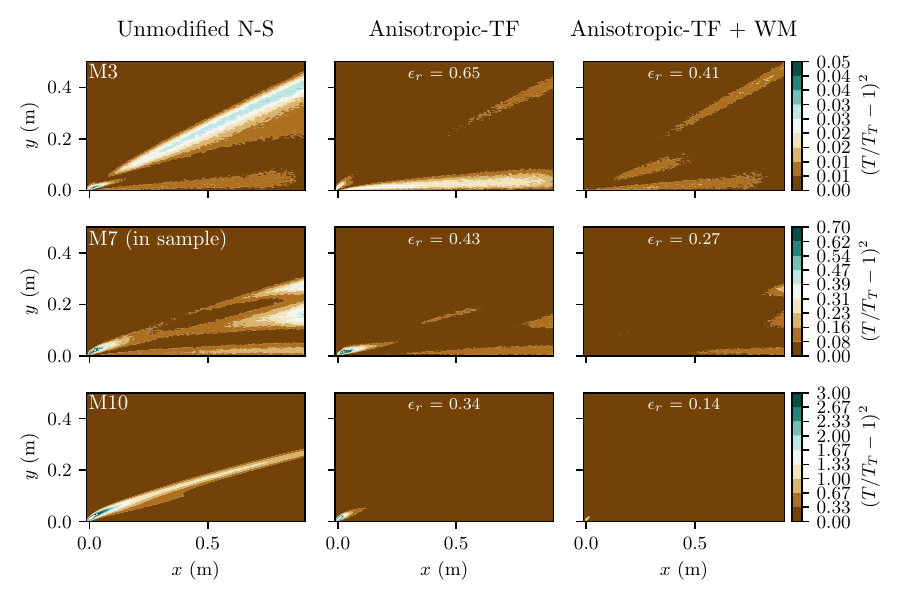} 
    \caption{Relative temperature errors  with and without the ML wall model (WM)  with respect to DSMC targets.}
    \label{fig: BW_Temp_Compare}
\end{figure}

Figures~\ref{fig: BW_Dens_Compare}, \ref{fig: BW_UVel_Compare}, and \ref{fig: BW_Temp_Compare} compare the unmodified Navier--Stokes density, streamwise velocity, and temperature predictions to Anisotropic-TF-augmented predictions with and without the ML wall model (WM).  Incorporating the wall model  reduces the density error for the in-sample $M_\infty=7$ case and the out-of-sample $M_\infty=10$ case; however, the improvements to the streamwise velocity and temperature are much more significant, as could be anticipated by the direct influence of the boundary conditions on these fields.
The wall model consistently reduces error in the streamwise velocity field by approximately 12\% for the in- and out-of-sample cases, with similar improvements observed for the predicted temperature field.  The most substantial improvements occur within the boundary layer, which is expected due to the direct influence of the boundary conditions in this region.

\begin{figure}
    \centering
    \includegraphics[width=0.49\textwidth]{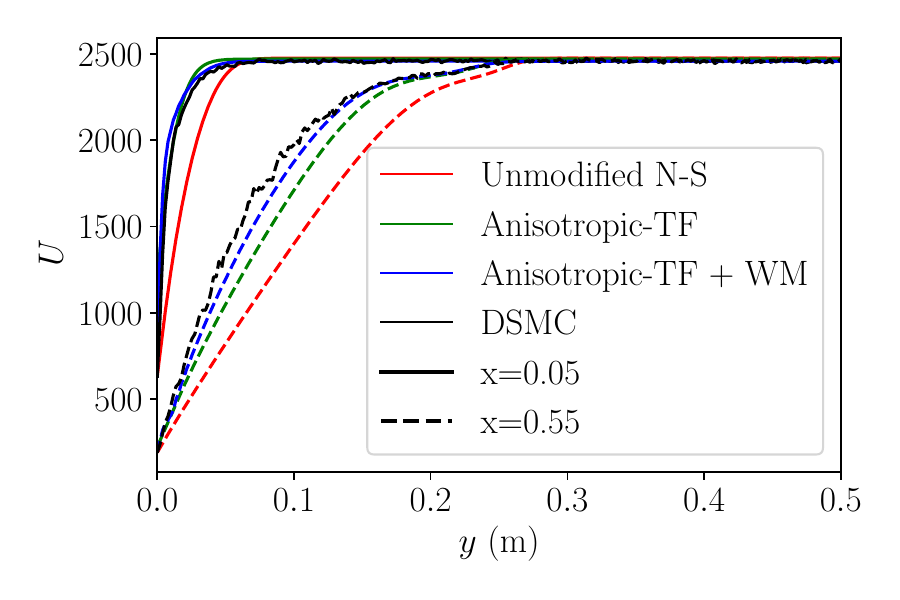} 
    \includegraphics[width=0.49\textwidth]{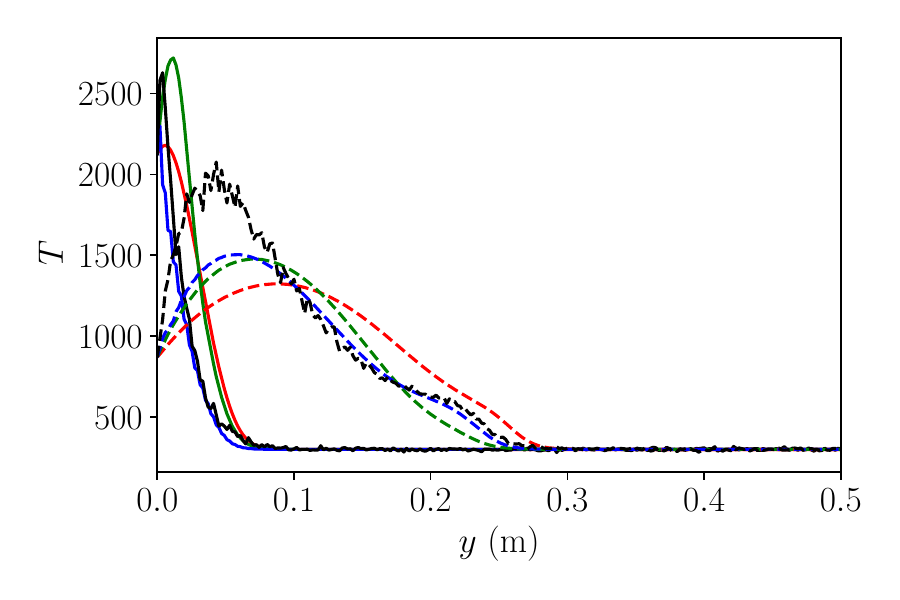} 
    \caption{Streamwise velocity and temperature predictions in the wall-normal direction at two points on the wall at ($x=0.05$ m and $x=0.55$ m) for the in-sample $M_\infty=7$, $\mathrm{Kn}=0.3$  case  with and without the ML wall model (WM).}
    \label{fig: BW_U_T_prof_compare}
\end{figure}

Figure~\ref{fig: BW_U_T_prof_compare} compares the wall-normal streamwise velocity and temperature profiles at streamwise locations \(x = 0.05\)~m and \(x = 0.55\)~m. The inclusion of the wall model leads to a substantial improvement in the streamwise velocity profile at \(x = 0.05\) m, with the predicted profile very closely matching the DSMC reference. At \(x = 0.55\) m, the wall-modeled prediction is also significantly closer to the DSMC profile compared to the prediction without the ML wall model. Notably, the addition of the wall model results in a velocity gradient near the wall that closely matches the DSMC profile, which is  important  for the viscous stress prediction, as we discuss subsequently.
The wall model likewise improves the wall-normal temperature profile, especially at \(x = 0.05\)~m. While the peak temperatures remain underpredicted, the wall model improves the wall-normal temperature gradient and hence the predicted wall-normal heat flux.

\subsubsection{Viscous Stress and Heat Flux}

\begin{figure}
    \centering
    \includegraphics[width=0.49\textwidth]{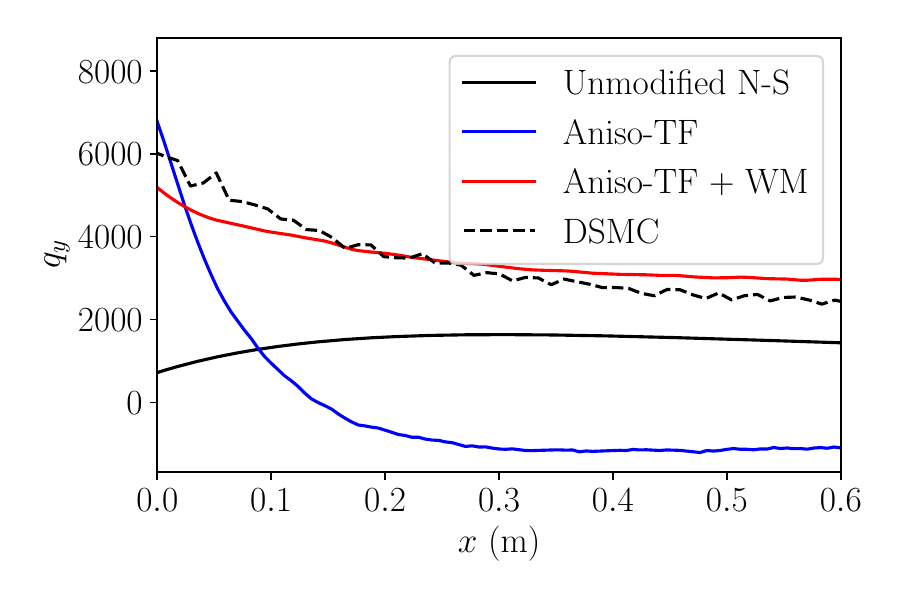} 
    \includegraphics[width=0.49\textwidth]{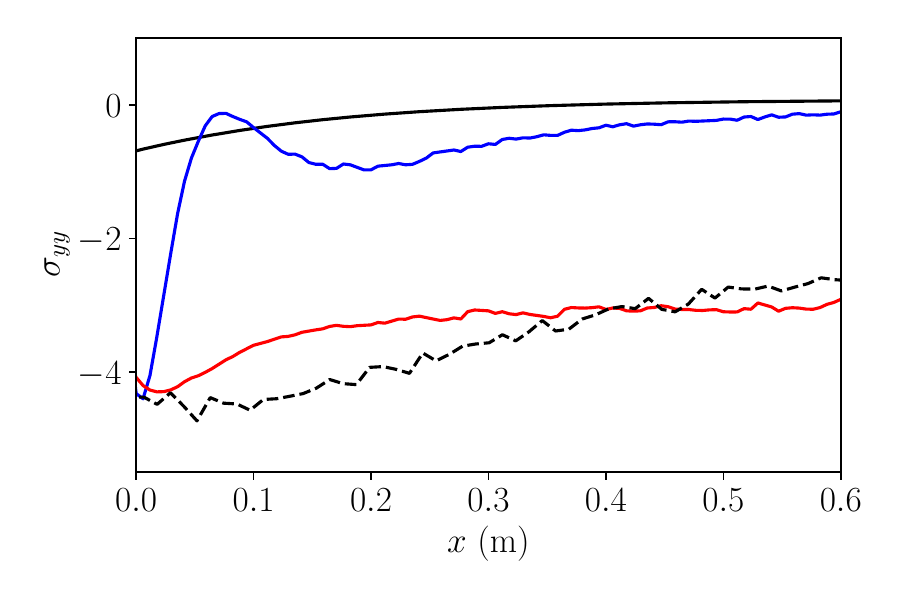} 
    \caption{Wall-normal heat flux and viscous stress along the wall for the in-sample $M_\infty=7$, $\mathrm{Kn}=0.3$ case.}
    \label{fig: BW_sigma_q_Compare}
\end{figure}

Figure~\ref{fig: BW_sigma_q_Compare} shows the predicted wall-normal  heat flux and viscous stress using the unmodified Navier--Stokes equations, Anisotropic-TF-augmented prediction without the wall model, Anisotropic-TF-augmented prediction with the wall model, and the target DSMC data for the in-sample $M_\infty=7$, $\mathrm{Kn}=0.3$ case. Adding the wall model does not impair the peak predicted heat flux and viscous stress at the flat-plate leading edge ($x=0$~m) but does significantly improve the predicted heat flux and viscous stress downstream along the wall, with the predicted fluxes  closely matching the DSMC results. This improvement aligns with observations from the previous section, where the wall model was shown to enhance near-wall gradients.

\begin{figure}
    \centering
    \includegraphics[width=0.49\textwidth]{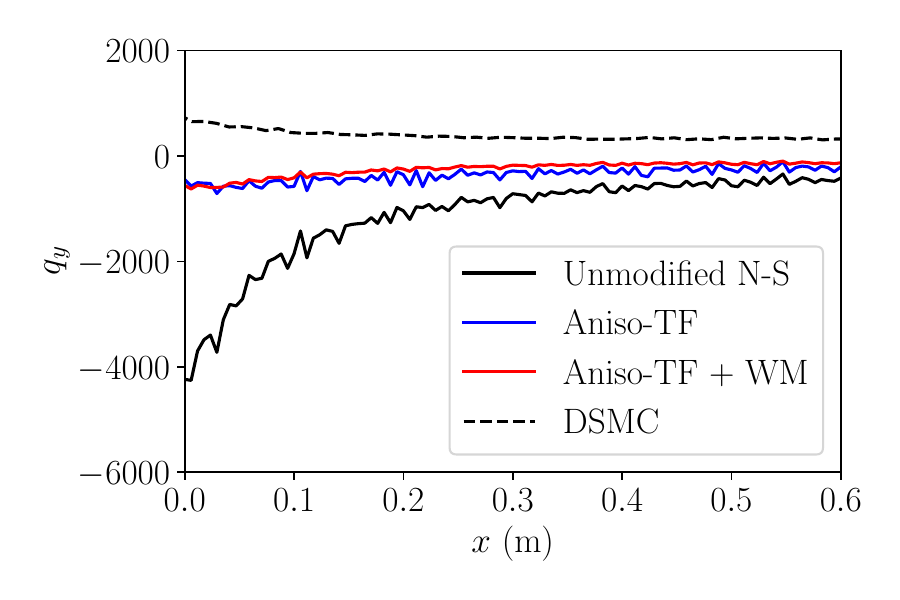} 
    \includegraphics[width=0.49\textwidth]{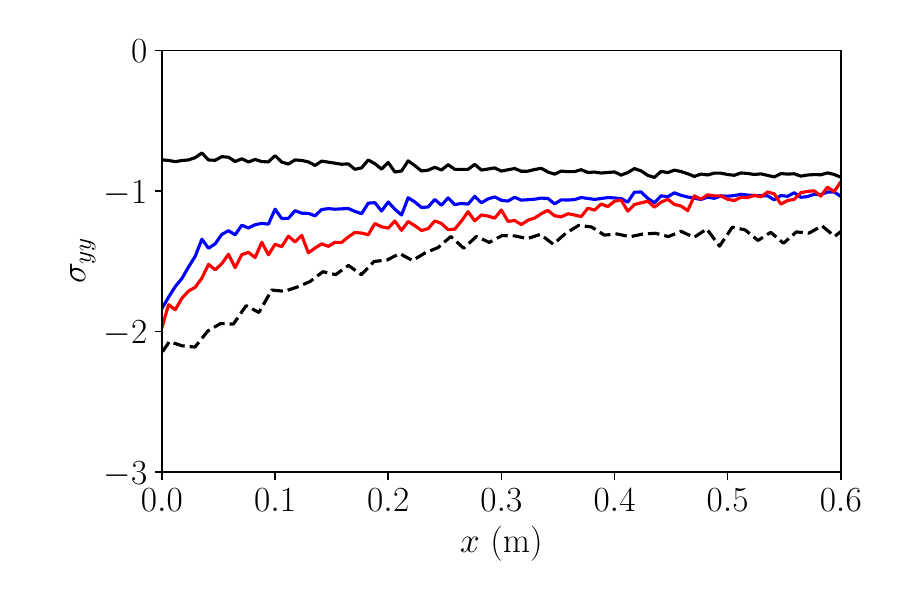} 
    \caption{Wall-normal heat flux and viscous stress along the wall for the out-of-sample $M_\infty=3$, $\mathrm{Kn}=0.3$ case. The ML models are trained for $M_\infty=7$, $\mathrm{Kn}=0.3$.}
    \label{fig: BW_M3_sigma_q_Compare}
\end{figure}

\begin{figure}
    \centering
    \includegraphics[width=0.49\textwidth]{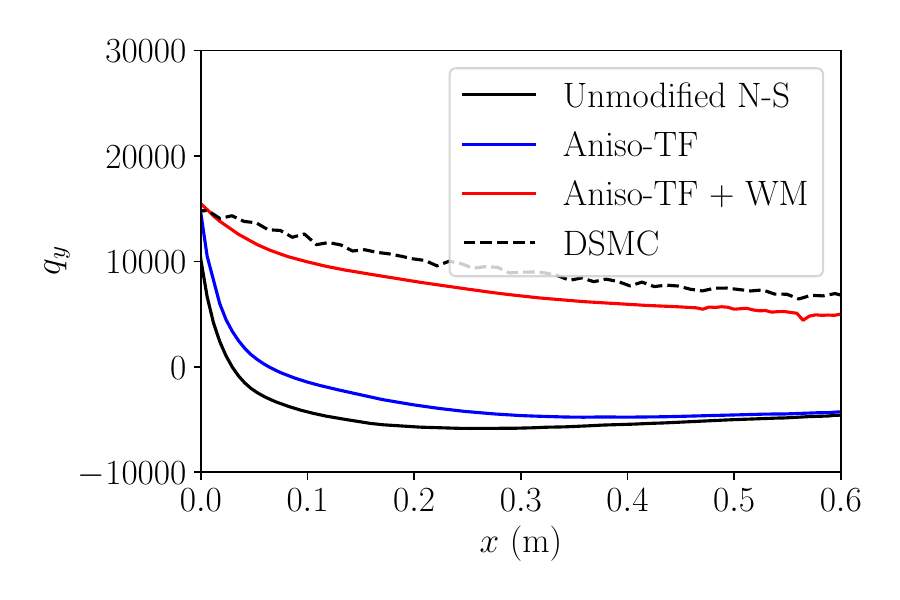} 
    \includegraphics[width=0.49\textwidth]{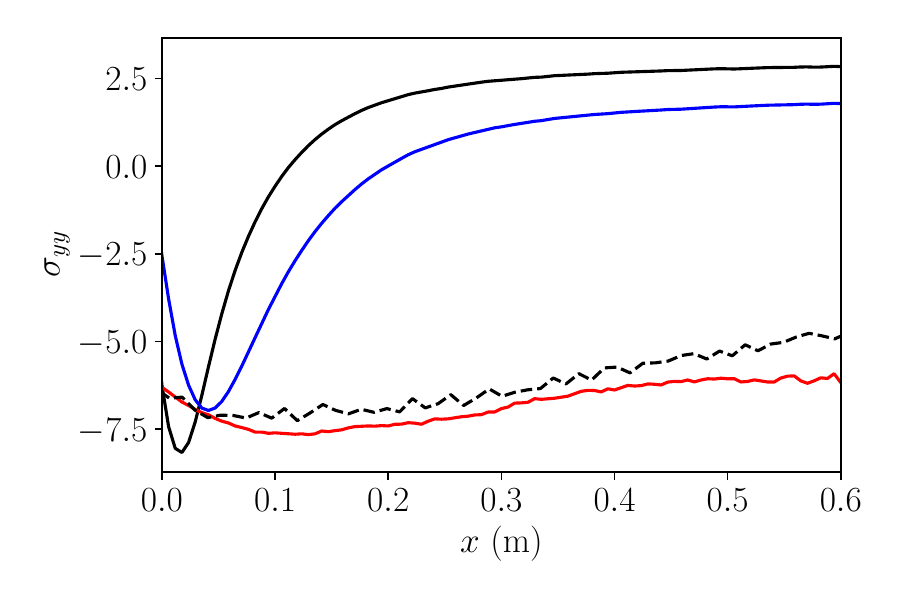} 
    \caption{Wall-normal heat flux and viscous stress along the wall for the out-of-sample $M_\infty=10$, $\mathrm{Kn}=0.3$ case. The ML models are trained for $M_\infty=7$, $\mathrm{Kn}=0.3$.}
    \label{fig: BW_M10_sigma_q_Compare}
\end{figure}

Figures~\ref{fig: BW_M3_sigma_q_Compare} and~\ref{fig: BW_M10_sigma_q_Compare} show analogous comparisons, with and without the ML wall model, for out-of-sample $M_\infty=3$ and $M_\infty=7$, $\mathrm{Kn}=0.3$ cases. For both, the unmodified Navier--Stokes solutions are qualitatively incorrect, particularly at $M_\infty=10$, for which the predicted heat flux and viscous stress have qualitatively incorrect signs. The ML wall model-augmented $M_\infty=3$ predictions are slightly improved, though the Anisotropic-TF model (without the wall model) already qualitatively improves the predictions at this relatively low out-of-sample Mach number.
Conversely, for $M_\infty=10$, the Anisotropic-TF model with the ML wall model succeeds in capturing the wall-normal heat flux and viscous stress along the entire boundary layer, yielding predicted profiles that closely align with the DSMC results, even for this out-of-sample Mach number.

\subsubsection{Extrapolation to Out-of-Sample Mach and Knudsen Numbers}

The previous sections demonstrated that both the ML transport coefficient models and the  ML wall model significantly improve solution accuracy for out-of-sample Mach numbers. Notably, the wall model leads to a significant reduction of error in the predicted wall-normal viscous stress and heat flux. We now evaluate the extrapolation capabilities of these models across across a range of transition-continuum Mach and Knudsen numbers.

All models discussed thus far have been trained for $M_\infty=7$ and a maximum local Knudsen number of approximately 0.3. We now consider test cases spanning $M_\infty\in\{2,3,7,9,10,12\}$ and  $\Kn = \{0.3,0.6,1.2\}$, for each of which we compare unmodified and ML-augmented Navier--Stokes predictions to DSMC-evaluated fields. We use a domain integrated relative $L_2$ error of the streamwise velocity,  temperature, and density,
\begin{equation}
  |\phi|_2^m = \frac{1}{3 A} \int_{\Omega} \left( |u|_2^N+ |T|_2^N + |\rho|_2^N \right) d \textbf{x},
  \label{eq:domain_L2}
\end{equation}
where $A$ is the  domain area {and $|u|_2^N$, $|T|_2^N$, $|\rho|_2^N$ are the normalized $L_2$ errors (\ref{eq: norm_L2}) for the streamwise velocity, temperature, and density}.

\begin{figure}
    \centering
    \includegraphics[width=0.49\textwidth, trim={0 30 0 20}, clip]{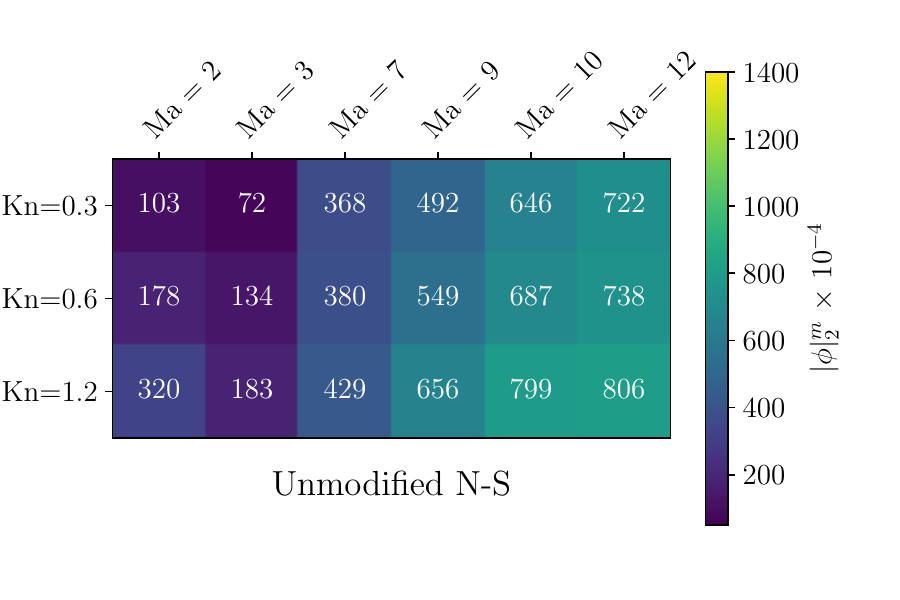} \\
    \includegraphics[width=0.49\textwidth, trim={0 30 0 20}, clip]{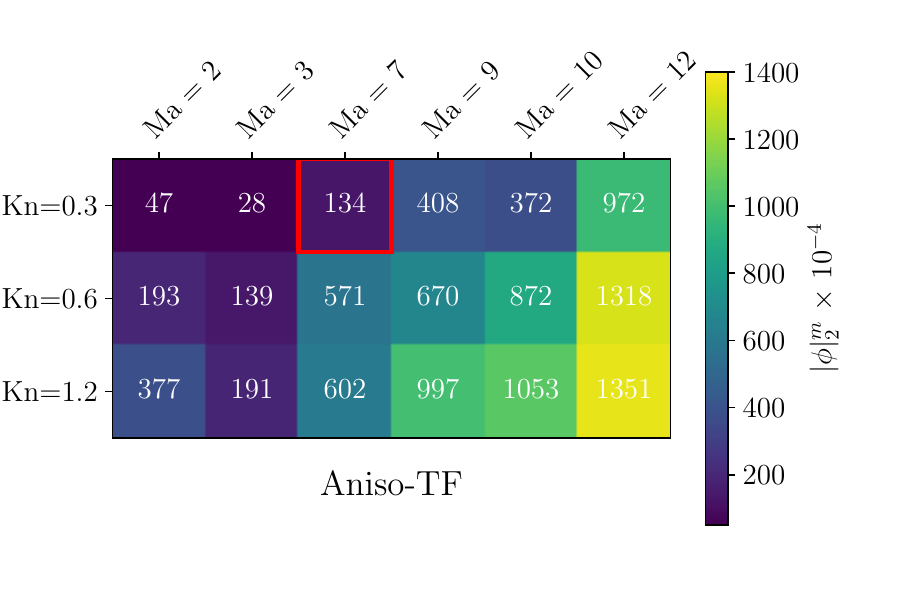} 
    \includegraphics[width=0.49\textwidth, trim={0 30 0 20}, clip]{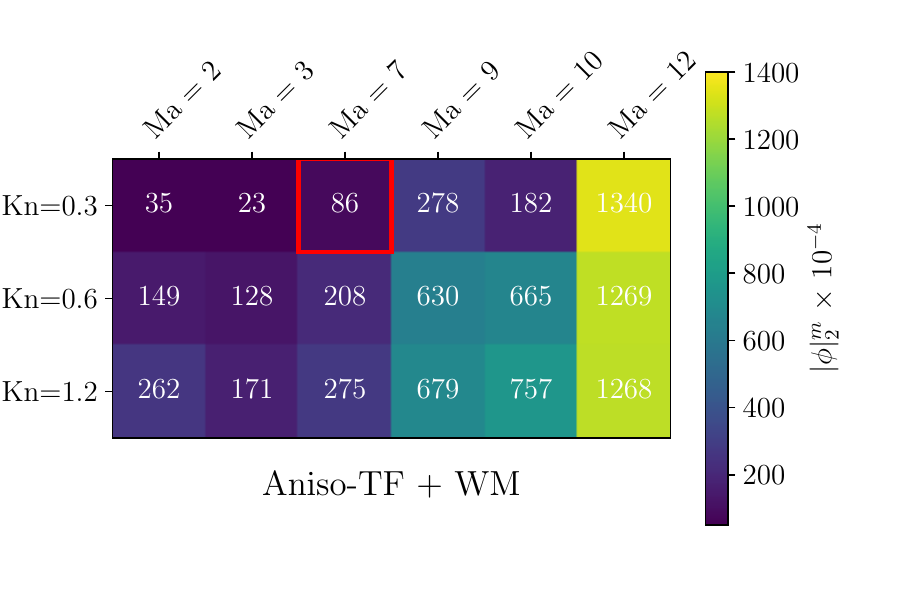} 
    \caption{Domain-integrated relative $L_2$ error \eqref{eq:domain_L2} for different in- and out-of-sample cases. Red-highlighted boxes indicate in-sample training cases.}
    \label{fig: BW_heatmaps}
\end{figure}

Figure~\ref{fig: BW_heatmaps}  shows the domain-integrated relative $L_2$ error of the unmodified Navier--Stokes prediction, the Anisotropic-TF-augmented predictions without the ML wall model, and the Anisotropic-TF-augmented predictions with the ML wall model.
As predicted by continuum theory limitations, the unmodified Navier--Stokes solution exhibits increasing $|\phi|_2^m$ with higher Mach and Knudsen numbers. This trend reflects the growing inaccuracy of the Navier–Stokes equations and the slip-wall boundary condition in more rarefied regimes (i.e., higher $\Kn$). The Anisotropic-TF transport model generally reduces the error relative to the baseline Navier--Stokes solution for out-of-sample Mach numbers, but only within the in-sample Knudsen number range ($\Kn = 0.3$). Notably, this improvement holds up to Mach 12, beyond which the model's extrapolation capability appears to break down.
However, when tested for out-of-sample Knudsen numbers ($\Kn = 0.6$ and 1.2), the transport models trained only for $\Kn=0.3$ are less reliable, in some cases yielding higher errors than the unmodified Navier--Stokes solution. These results indicate that while the transport models enhance performance near their Knudsen number training condition, on its own the single-Knudsen number training does not guarantee extrapolation to more rarefied conditions.

The co-trained ML wall model leads to additional improvements, particularly within the in-sample Knudsen number regime. The heatmaps show consistent reductions in $|\phi|_2^m$ relative to the transport model alone. However, the wall model also fails to maintain accuracy when extrapolated to Mach 12. For higher Knudsen numbers, the wall model offers marginal improvements over the baseline Navier--Stokes solution, except for the in-sample Mach 7 case, where it performs notably better. Despite this, the co-trained wall model consistently outperforms the transport model alone across the entire regime map.

\subsubsection{Parallel Training}

To improve predictive performance across different $\Kn$, we adopt a parallel training strategy in which the model is trained concurrently over three Knudsen numbers $\Kn = [0.3,0.6,1.2]$. This is achieved by averaging the gradients from each parallel process before applying the parameter update during training, as described in Section~\ref{sec:parallel_training}.

\begin{figure}
    \centering
    \includegraphics[width=0.49\textwidth, trim={0 30 0 20}, clip]{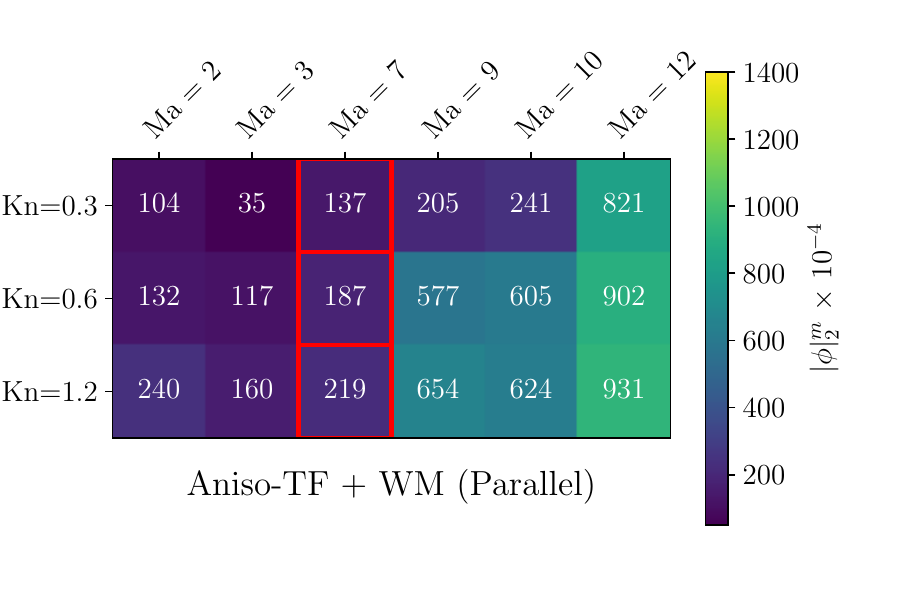} 
    \includegraphics[width=0.49\textwidth, trim={0 30 0 20}, clip]{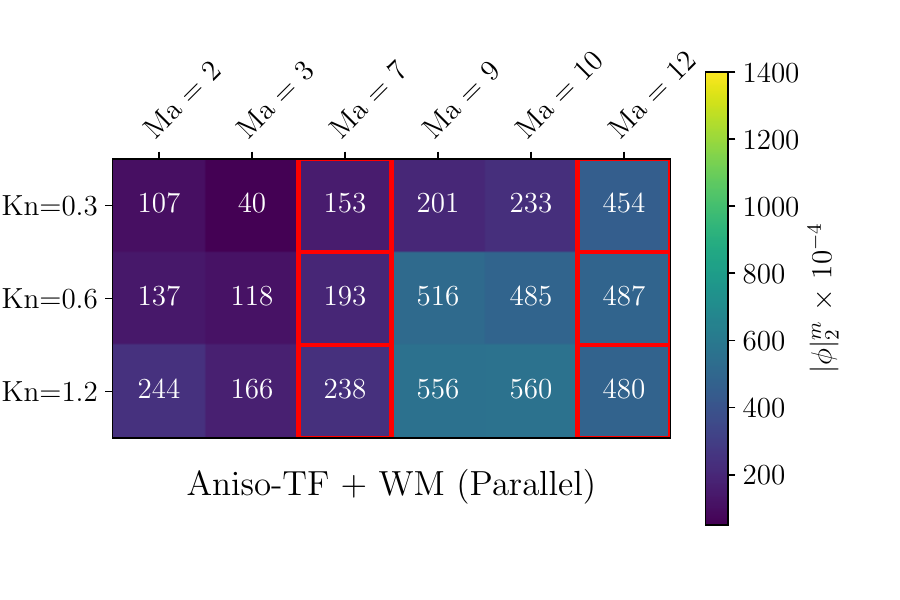}
    \caption{Domain-integrated relative \(L_2\) error \eqref{eq:domain_L2} for multi-condition-trained models. {Red-highlighted boxes indicate in-sample training cases.}
    }
    \label{fig: BW_heatmap_6Kn}
\end{figure}

Figure~\ref{fig: BW_heatmap_6Kn} shows the domain-integrated error for parallel-trained models trained for $M_\infty=7$ and $M_\infty=[7,12]$. The $M_\infty=7$-trained model shows a marked improvement in accuracy compared to the unmodified Navier--Stokes solution across all Mach numbers less than 12 and across the Knudsen number range. Although there is a slight increase in error for  $\Kn = 0.3$ compared to the model trained only on this Knudsen number, this is anticipated as the $\Kn = [0.3,0.6,1.2]$-trained model is optimized to minimize the error summed over multiple conditions.

Since training over three Knudsen numbers ($\Kn = [0.3,0.6,1.2]$) does not improve performance for $M_\infty = 12$ case, we next explicitly include $M_\infty=12$ in the training cases across the same three $\Kn$ regimes in the training dataset. Figure~\ref{fig: BW_heatmap_6Kn} shows the mean relative error for this extended training configuration, for which including the $M_\infty=12$ data in training  significantly reduces the prediction error across the entire regime map. This confirms that the model benefits from direct exposure to high-Mach, nonequilibrium flow data. Notably, this model also exhibits improved performance at intermediate Mach numbers, particularly Mach 9 and Mach 10, which were not seen during training. This demonstrates that incorporating the full range of Mach number extremes (Mach 7 and Mach 12) enables better interpolation across the Mach number spectrum. While a slight increase in error is observed for the lower Mach number cases (e.g., Mach 3 and Mach 5), the overall performance across the full Mach–Knudsen number space is the most balanced among all the models considered. By including training data from both ends of the Mach spectrum, this model achieves the best generalization capability across all tested Mach and Knudsen number conditions.

\begin{figure}
    \centering
    \includegraphics[width=0.50\textwidth]{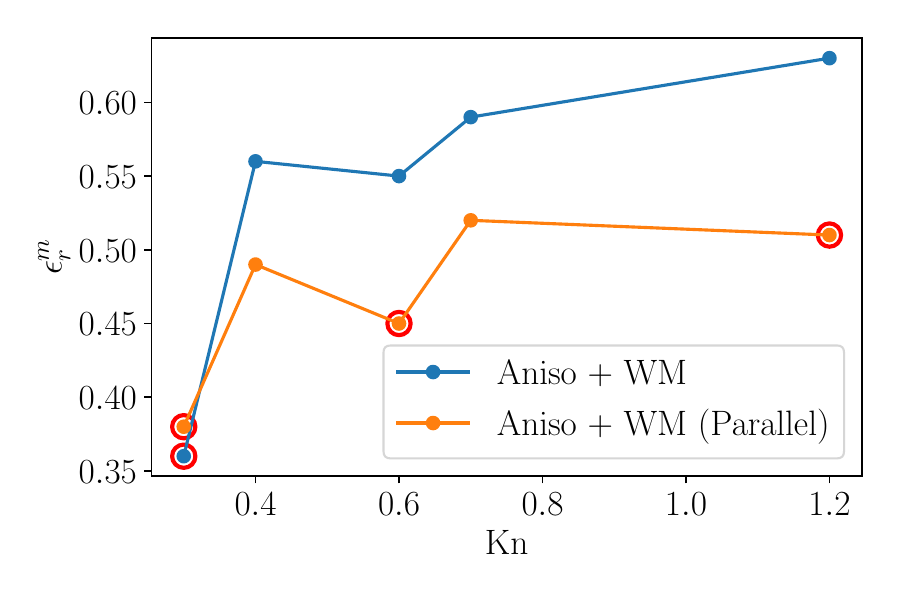}
    \vspace{-0.5em}
    \caption{Mean relative error \(\epsilon_{r}^m\) across Knudsen numbers for single-\(\Kn\) and parallel-trained models. Red circles indicate in-sample training cases. }
    \label{fig: KnInterp}
\end{figure}

To further evaluate the effectiveness of the parallel training strategy, we define a relative error metric averaged over the density, temperature, and streamwise velocity, 
\begin{equation} 
    \epsilon_{r}^m = \left( \epsilon_{r,\rho} + \epsilon_{r,T} + \epsilon_{r,u} \right).
\end{equation}
Figure~\ref{fig: KnInterp} presents $\epsilon_{r}^m$ for models trained on a single Knudsen number and for those trained using the parallel training approach and tested for a finer range of Knudsen numbers than is shown in Figures~\ref{fig: BW_heatmaps} and \ref{fig: BW_heatmap_6Kn}. As expected, the model trained for a single $\Kn$ exhibits increasing error for Knudsen numbers away from its training regime. Conversely, the parallel-trained model not only achieves lower error for its in-sample Knudsen numbers but also maintains relatively consistent performance across the entire testing range, even for out-of-sample cases. These results suggest that parallel training over multiple Knudsen numbers improves the model's generalization capability across the regime map.

\subsubsection{Parameter Convergence Study}

We next investigate the influence of increasing the {degrees of freedom} of the neural network closures used for the transport and wall models. All previously tested models utilized neural networks with 30 neurons in each hidden layer. In this analysis, we evaluate the impact of using 100 neurons per layer, thereby significantly increasing the flexibility of the model’s functional form.

\begin{figure}
    \centering
    \includegraphics[width=0.55\textwidth, trim={0 30 0 20}, clip]{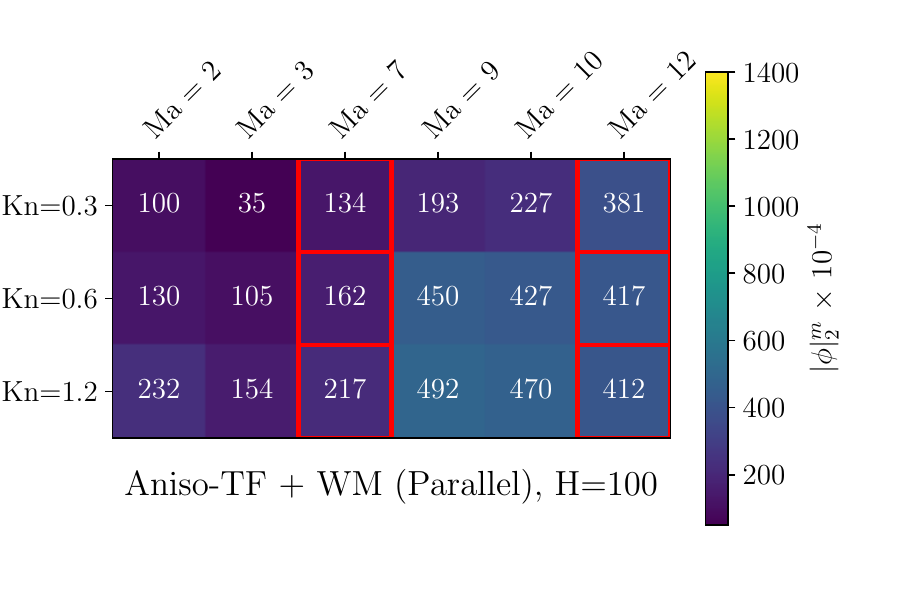} 
    \vspace{-0.5em}
    \caption{Domain-integrated relative \(L_2\) error \eqref{eq:domain_L2} for a multi-condition-trained model with $H=100$ neurons per hidden layer. {Red-highlighted boxes indicate in-sample training cases.}}
    \label{fig: BW_heatmap_6Kn_H100}
\end{figure}

Figure~\ref{fig: BW_heatmap_6Kn_H100} shows the mean relative error $|\phi|_2^m$ for models trained with 100 neurons per layer using the parallel training approach. Compared to the analogously trained model in Figure~\ref{fig: BW_heatmap_6Kn}, this increase in {degrees of freedom} provides a notable reduction of error for in-sample cases but does not fundamentally change the overall out-of-sample extrapolation trends. While improvements are generally observed for out-of-sample cases, these do not scale proportionally with the more than threefold increase in the number of neural network parameters. This suggests diminishing returns with respect to neural network {degrees of freedom} and highlights the potential risk of overfitting to in-sample cases, thereby compromising generalizability across the full range of Mach and Knudsen numbers.

It is also important to note that increasing the number of neurons from 30 to 100 comes with a substantial increase in computational cost during training. These findings underscore the trade-off between model {degrees of freedom}, generalization performance, and training efficiency.

\subsection{Cost Comparison}

The main reason for simulating transition-continuum flows using the Navier--Stokes equations is their potential for cost reduction compared to DSMC.
We now assess the computational cost of ML-augmented Navier--Stokes solutions compared to that of the target DSMC predictions and unmodified Navier--Stokes solutions. We consider Anisotropic–TF-augmented Navier--Stokes predictions with the ML wall model, which have the greatest evaluation cost of the aforementioned ML closures. 

We perform all simulations on an NVIDIA RTX A2000 GPU with 12 GB of memory. The Navier--Stokes solver (\emph{PyFlowCL} \cite{liu2024adjoint,hickling2024large}) performs all arithmetic operations on the GPU using NVIDIA's cuBLAS library via \emph{PyTorch} \cite{paszke2019pytorchimperativestylehighperformance}. We compile \emph{SPARTA} for GPU calculations using the \emph{Kokkos} framework \cite{9485033}, which accelerates operations on the GPU using the cuBLAS and cuFFT libraries. All calculations use 64-bit floating-point precision.

We advance the Navier--Stokes equations until the forward residuals converge to an absolute error of $10^{-8}$. We advance the DSMC for 150,000 iterations, with two adaptive mesh refinement steps applied after 50,000 and 100,000 iterations. We employ value-based, $h$-adaptive refinement and coarsening on the Cartesian DSMC grid using \emph{SPARTA}’s $\lambda/\Delta$ metric, refining cells where $\lambda/\Delta \le 2$ and coarsening otherwise, to accurately resolve Knudsen layers and shock gradients while controlling statistical noise. 
 
Table~\ref{tab:cost_comp} summarizes the total wall-clock times for DSMC, unmodified Navier--Stokes, and ML-augmented Navier--Stokes simulations for $\Kn = 0.3$, $0.6$, and $1.2$ at $M_\infty = 7$. As expected, the DSMC cost is inversely proportional to the Knudsen number, since the number of computational particles and hence the overall cost scales in proportion to the gas density. 
Compared to DSMC, the unmodified Navier--Stokes simulations are between 6.5 times faster ($\Kn=1.2$) and  22 times faster ($\Kn=0.3$), though at concomitantly larger \emph{a posteriori} error (see Figure~\ref{fig: BW_heatmaps}).
Due to the need to evaluate the neural network models, the ML-augmented Navier--Stokes simulations incur 33\,\%  cost penalty relative to the baseline unmodified Navier--Stokes predictions. Despite this increased cost, the ML-augmented Navier--Stokes predictions nonetheless accelerate predictions by factors of 4.9  ($\Kn=1.2$) to 15.6 ($\Kn=0.3$) compared to DSMC.

In view of their substantial improvement in predictive accuracy, evaluating the ML transport and wall models introduces a relatively marginal increase in cost over the baseline Navier--Stokes predictions while remaining significantly more economical than DSMC across the present range of Knudsen numbers. The attractiveness of ML-augmented Navier--Stokes predictions of course increases as the Knudsen number decreases, in which regime the ML-augmented continuum predictions become increasingly accurate and DSMC solutions are substantially more expensive.

\begin{table}
\centering
\caption{Walltime cost of DSMC, unmodified Navier--Stokes, and ML-augmented Navier--Stokes predictions, rounded to the nearest minute, using an NVIDIA RTX A2000 GPU. All simulations are for $M_\infty = 7$. Speedup factors are given with respect to the DSMC cost.}
\label{tab:cost_comp}
\begin{tabular}{l | c | c c | c c}
\toprule
 & DSMC  & \multicolumn{2}{c|}{Unmodified NS} & \multicolumn{2}{c}{Anisotropic-TF + WM} \\
Kn & Walltime (H:MM) & Walltime (H:MM) & Speedup & Walltime (H:MM) & Speedup \\
\midrule
0.3 & 7:01 & 0:19 & 22.0 & 0:27 & 15.6 \\
0.6 & 3:50 & 0:20 & 11.4 & 0:26 & 8.9 \\
1.2 & 2:08 & 0:20 & 6.5 & 0:26 & 4.9 \\
\bottomrule
\end{tabular}
\end{table}

\section{Extension to out-of-sample geometries} \label{sec:out-of-sample-geo}

\begin{figure}
    \centering \includegraphics[width=0.55\textwidth]{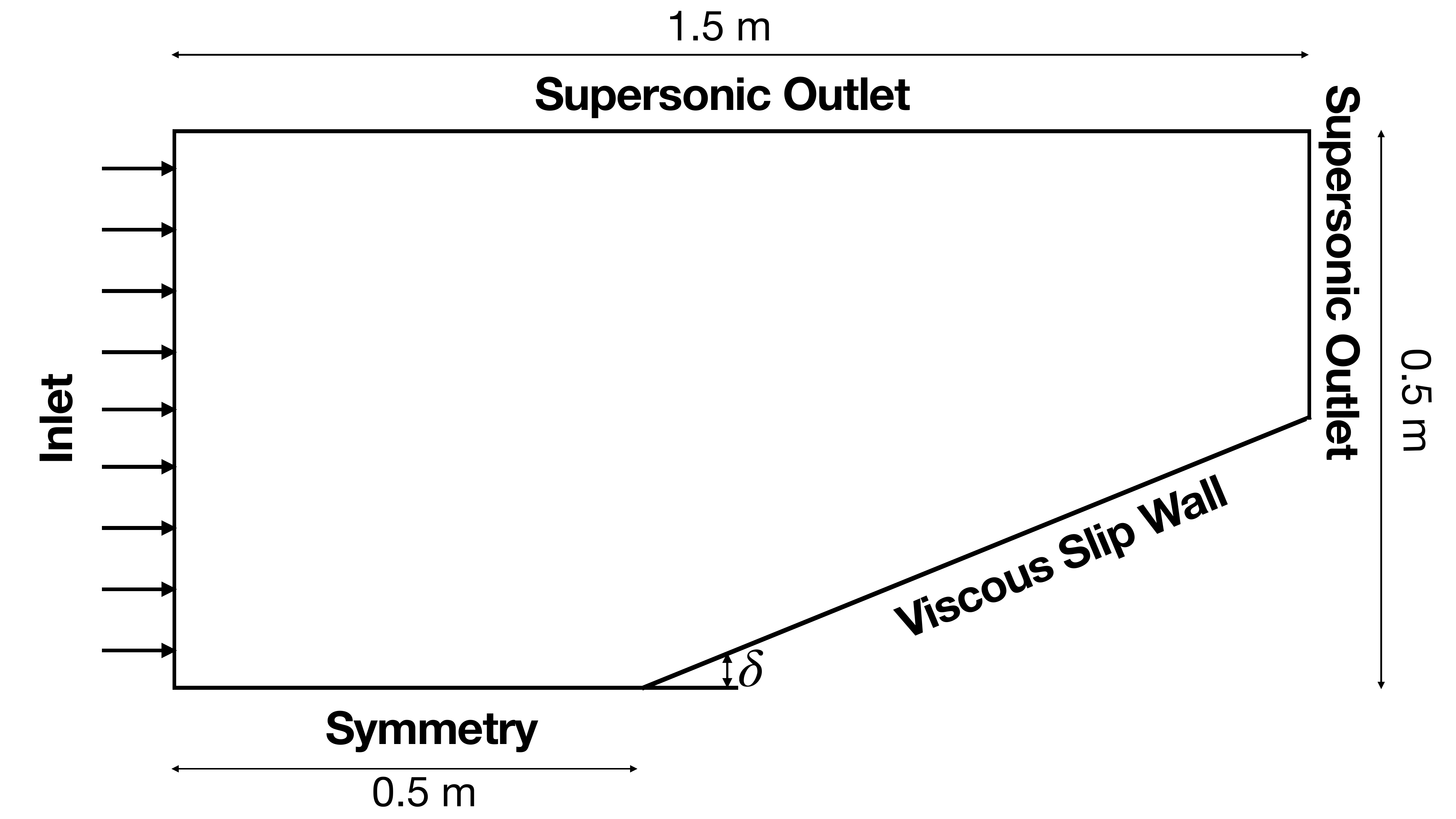} 
    \caption{Schematic of the inclined wedge geometry showing                                                                                                                   domain dimensions, boundary conditions and wedge angle $\delta$.
    } 
    \label{fig:InclinedWedgeSchem} 
\end{figure}

The combined ML transport coefficient model and wall model trained concurrently across three Knudsen numbers ($\Kn = [0.3,0.6,1.2]$) and two Mach numbers ($M_\infty = [7,12]$) demonstrated the strongest generalization within the sampled parameter space (see Figure~\ref{fig: BW_heatmap_6Kn}). To further assess this model's robustness, we evaluate its performance for an out-of-sample geometry---an inclined wedge, which exhibits flow redirection not present in the flat-plate training cases. Figure~\ref{fig:InclinedWedgeSchem} depicts the computational domain and boundary conditions for this case. The configuration closely parallels the previously studied flat plate, with the key distinction that the wall is inclined at an angle $\delta$ relative to the freestream direction. To test generalization across varying wall inclinations, we consider three wedge angles, $\delta = 2.5^\circ$, $5^\circ$ and $10^\circ$, for $\Kn=0.3$ and $M_\infty = 5$. We note that this Knudsen number is in-sample for this model, while the Mach number is out-of-sample (extrapolation).

\begin{figure}
    \centering
    \includegraphics[width=0.65\textwidth]{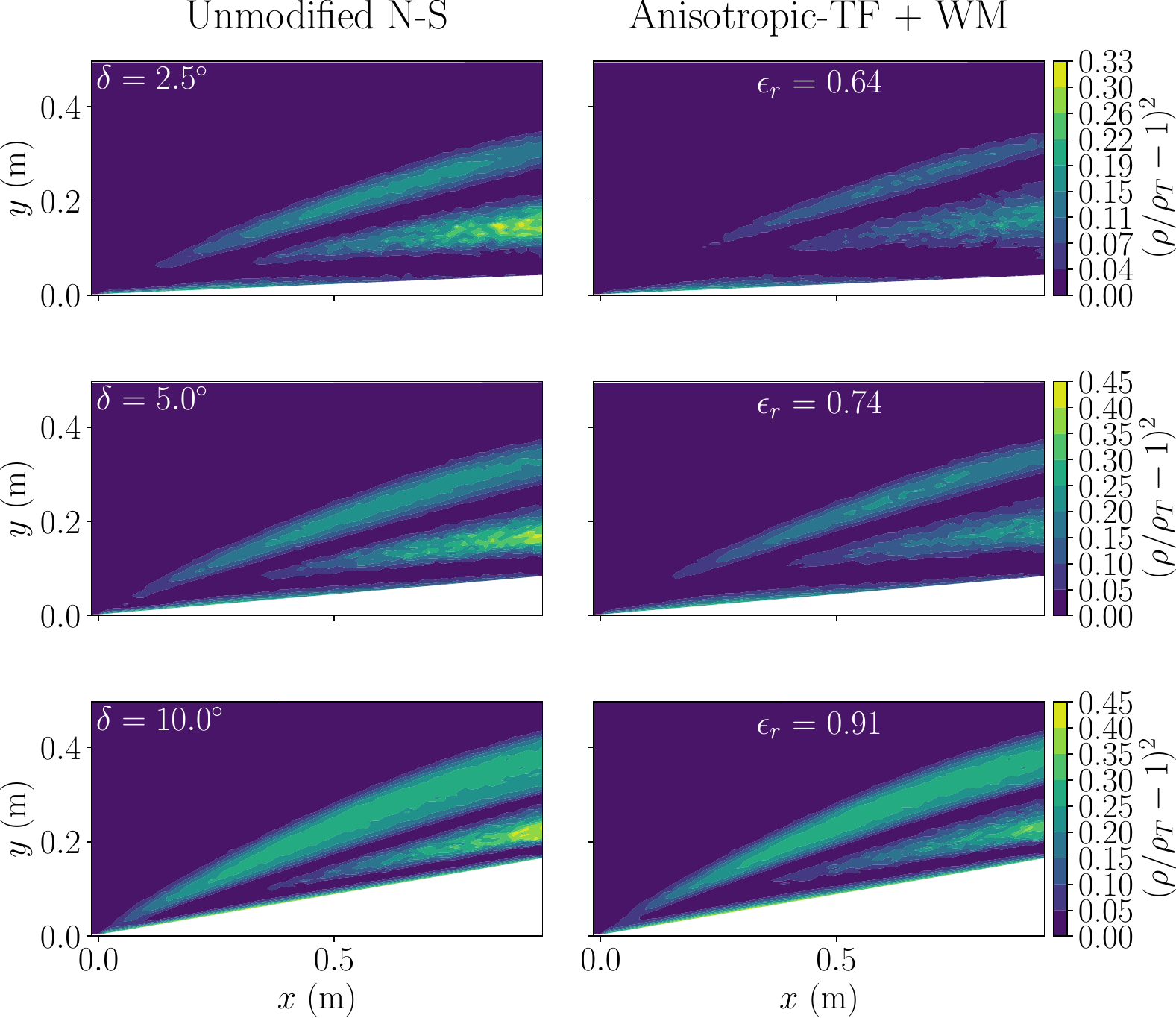} 
    \caption{Relative density errors for the inclined wedge configuration with respect to DSMC targets. $\epsilon_r$ is the domain L2 relative error.}
    \label{fig: Inc_wedge_Dens_Compare}
\end{figure}

\begin{figure}
    \centering
    \includegraphics[width=0.65\textwidth]{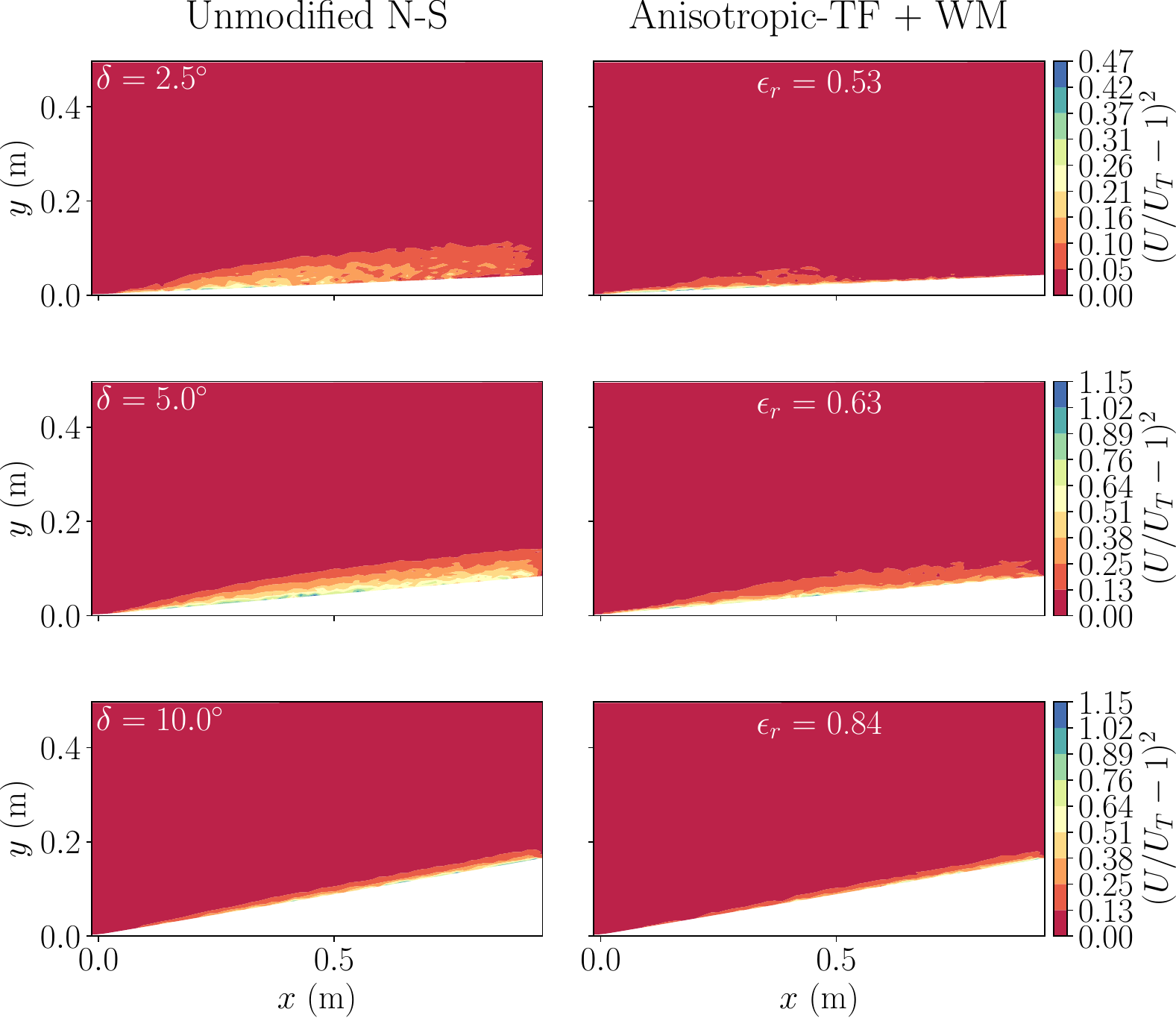}
    \caption{Relative streamwise velocity errors for the inclined wedge configuration with respect to DSMC targets. $\epsilon_r$ is the domain L2 relative error.}
    \label{fig: Inc_wedge_Vel_Compare}
\end{figure}

\begin{figure}
    \centering
    \includegraphics[width=0.65\textwidth]{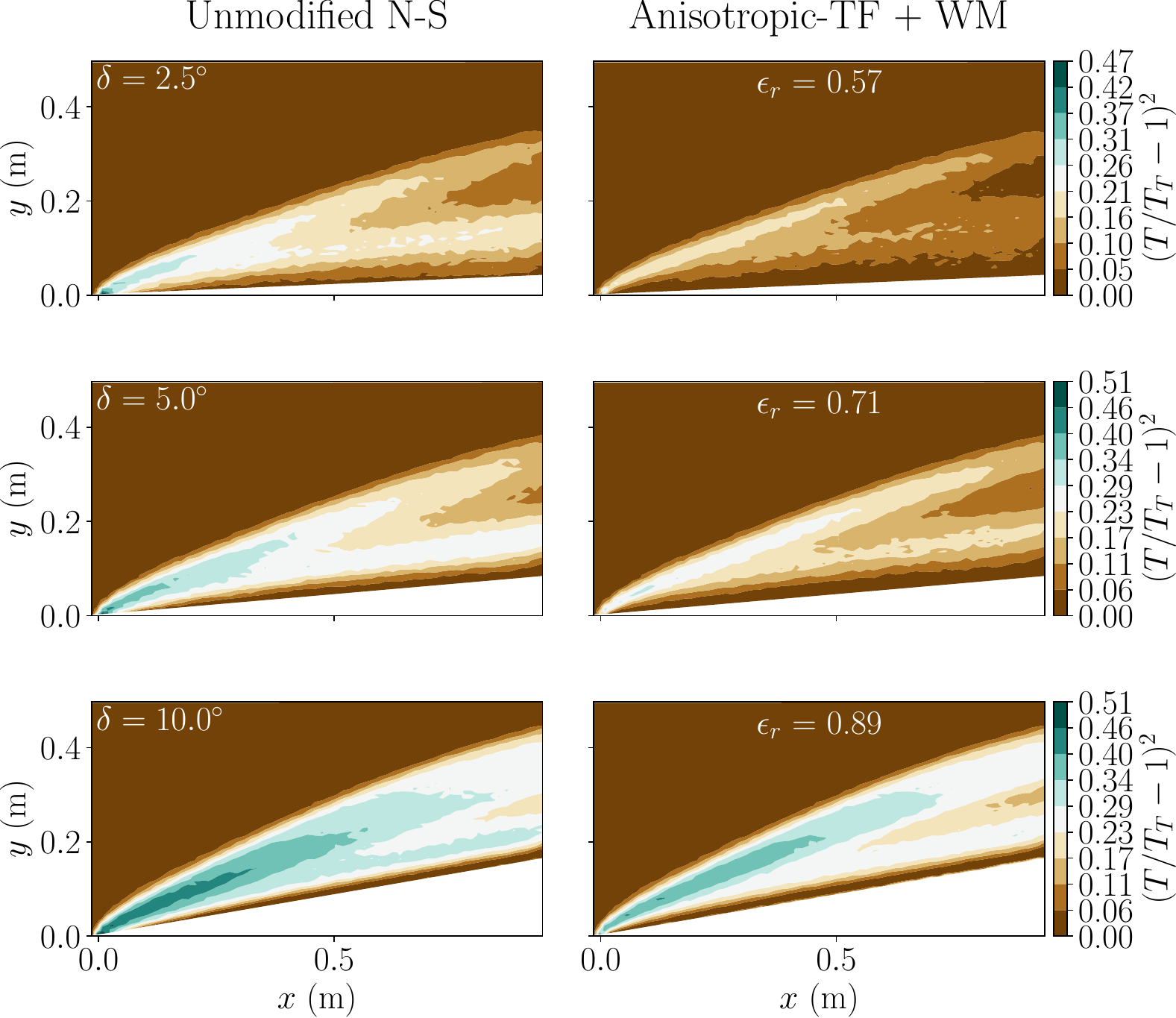} 
    \caption{Relative temperature errors for the inclined wedge configuration with respect to DSMC targets. $\epsilon_r$ is the domain L2 relative error.}
    \label{fig: Inc_wedge_Temp_Compare}
\end{figure}

Figures~\ref{fig: Inc_wedge_Dens_Compare}, \ref{fig: Inc_wedge_Vel_Compare}, and~\ref{fig: Inc_wedge_Temp_Compare} compare the unmodified Navier–Stokes predictions of density, streamwise velocity, and temperature to those obtained using the Anisotropic-TF transport model coupled with the wall model for all three wedge angles. The local relative error with respect to the DSMC ($\epsilon_r$) is reported in the figures for quantitative assessment. For all wedge angles, ML augmentation substantially improve the Navier--Stokes equations' predictive accuracy, even for this out-of-sample geometry.

For density, the augmented model improves $\epsilon_r$ by nearly 40\,\% at the $\delta=2.5^\circ$ wedge angle, with the improvement decreasing to about 30\,\% for $\delta=5^\circ$ and to 9\,\% for $\delta=10^\circ$. A similar trend is observed for the streamwise velocity field, for which the relative error improvement diminishes with increasing $\delta$. As seen in Figure~\ref{fig: Inc_wedge_Vel_Compare}, the boundary-layer error grows for larger inclinations, indicating that the wall model loses effectiveness as $\delta$ increases. The temperature fields show a comparable trend, with the improvement diminishing at larger wedge angles.

The monotonic increase of $\epsilon_r$ with wall angle is consistent with the expectation that the models' augmentation capability diminishes for increasingly out-of-sample geometries. The present model's training case (the flat plate) exhibits only minor flow redirection due to boundary layer growth. For increasing wedge angles, however, flow redirection and oblique-shock compression become more dominant, both of which are outside the training conditions and which reduce the model’s corrective abilities.
Notably, only marginal improvements are apparent within the shock layer, suggesting that the flat plate-trained transport model's augmentation capability diminishes more rapidly than that of the ML boundary model.

Nonetheless, it is clear that the flat plate-trained models generalize relatively well to modestly out-of-sample geometries. The fact that the approach provides stable solutions---let alone improved accuracy---is a feature of the PDE-constrained ML optimization method \cite{sirignano2020dpm,macart2021embedded}.  However, as could be expected for any data-driven model, the trained model's effectiveness diminishes for increasing deviations from the training data. This underscores the need for broader training datasets to capture dominant flow features for configurations of engineering interest.

\section{Conclusion} \label{sec:conclusion}

We present a {physics-based} machine learning framework for improving continuum-scale modeling of nonequilibrium hypersonic flows. The framework addresses two key modeling challenges for hypersonic wall-bounded flows: (i) nonequilibrium viscous and thermal-diffusive transport, and (ii) nonequilibrium wall boundary conditions. The approach embeds untrained neural networks into the Navier--Stokes equations and trains these networks via {adjoint optimization}, which ensures that the learned corrections remain consistent with the governing physical laws.  We develop and test closures for hypersonic transition-continuum boundary layers, focusing on a regime in which the Navier--Stokes formulations and existing, semi-empirical boundary conditions give inaccurate predictions compared to DSMC targets.

We propose and evaluate three neural network transport closures with different forms of the augmented viscosity and thermal conductivity tensors: isotropic, anisotropic, and anisotropic trace-free. We train these models for $M_\infty=7$ and domain-maximum Knudsen number $\Kn=0.3$ and evaluate them for a range of out-of-sample Mach and Knudsen numbers. The anisotropic trace-free model, which enforces key invariance properties of the viscous stress tensor and heat flux vector, outperforms the simpler closures for in- and out-of-sample conditions.

We additionally develop hypersonic wall boundary-condition models based on modeled non-Maxwellian distribution functions. These offer a physics-driven alternative to the conventional Maxwellian slip and temperature jump boundary conditions. By modeling the near-wall velocity distribution as a convex combination of skewed Gaussians, the non-Maxwellian wall model enables accurate predictions of the near-wall heat flux and viscous stress, especially in rarefied regimes having highly bimodal velocity distributions. The non-Maxwellian boundary conditions, adjoint-optimized simultaneously with the neural network transport closures, substantially improve near-wall predictions and alleviate the limitations of the Maxwellian slip models at higher Mach and Knudsen numbers.

Extensions of the combined transport and wall models are trained for multiple Mach and Knudsen numbers, which significantly improves the models' interpolation and extrapolation capability for $M_\infty\in[2,12]$ and  $\Kn\in[0.3,1.2]$.
The parallel-training strategy we present could be readily extended to train models for still-larger parameter spaces. Significantly, the multiple-condition-trained models improve predictive accuracy of the augmented Navier--Stokes equation across the entire regime map that we test.
Increasing the network size (number of neural network degrees of freedom) does improve the models' predictive accuracy overall, though the accuracy gains are not commensurate with the increased complexity of the network, indicating that the network size is sufficient for the modeling task.
The ML-augmented Navier--Stokes simulations are  five to fifteen times faster than DSMC, depending on the Knudsen number, confirming that the proposed framework can accelerate transition-continuum predictions while preserving physical fidelity.

We further evaluate the model’s generalization capability on an out-of-sample inclined wedge geometry for wall angles between $2.5^\circ$ and $10^\circ$. The ML-augmented model reproduces key flow features and trends in the nonequilibrium regions even for this out-of-sample flow, though its accuracy decreases with increasing wedge angle, which is consistent with the geometric deviation from model’s flat-plate training configuration. The model's stability and accuracy for these geometrically out-of-sample flows are encouraging and are consistent with previous studies of PDE-constrained ML optimization. Nonetheless, for engineering applications, it is essential to include geometrically representative training configurations---as is true for any data-driven model.

This work demonstrates the promising capability of PDE-embedded machine learning closure models, trained using physics-constrained, adjoint-based optimization,  to enhance the predictive accuracy of continuum solvers for hypersonic nonequilibrium flows. By replacing empirical closure assumptions with data-driven, physically grounded models, the physics-based machine learning framework  extends the predictive capability of the Navier--Stokes equations deeper into the transition-continuum regime.
Ongoing  efforts focus on extending this framework to additional nonequilibrium phenomena, such as rotational and vibrational excitation and multicomponent and chemically reacting flows, higher-order continuum transport closures, and geometrically more complicated flows.

\section*{Acknowledgments}
This material is based upon work supported by the Department of Defense, Office of Naval Research, under Award Number N00014-22-1-2441. The authors gratefully acknowledge computing resources provided by the Center for Research Computing at the University of Notre Dame. 

\section*{Data Availability}
The computer codes and datasets required to reproduce the findings contained herein will be made available at \url{https://doi.org/10.5281/zenodo.15678863} \cite{dataset}.


\begin{thebibliography}{10}
\expandafter\ifx\csname url\endcsname\relax
  \def\url#1{\texttt{#1}}\fi
\providecommand{\href}[2]{#2}
\providecommand{\path}[1]{#1}
\expandafter\ifx\csname urlprefix\endcsname\relax\def\urlprefix{URL }\fi
\expandafter\ifx\csname href\endcsname\relax
  \def\href#1#2{#2} \def\path#1{#1}\fi
\providecommand{\DOIprefix}{doi:}
\providecommand{\doi}[1]{\href{http://dx.doi.org/#1}{\path{#1}}}

\bibitem{JJShuFluidVelocitySlip}
J.-J. Shu, J.~Bin Melvin~Teo, W.~Kong~Chan, Fluid Velocity Slip and Temperature
  Jump at a Solid Surface, \emph{Applied Mechanics Reviews} 69~(2) (2017)
  020801. doi:\doi{10.1115/1.4036191}.
\newblock

\bibitem{Vincenti1917}
W.~G. Vincenti, C.~H. Kruger, \emph{Introduction to Physical Gas Dynamics}, New
  York, Wiley, 1965.

\bibitem{Greenshields2007}
C.~J. Greenshields, J.~M. Resse, The structure of shock waves as a test of
  Brenner’s modifications to the Navier–Stokes equations, \emph{Journal of
  Fluid Mechanics} 580 (2007) 407–429. doi:\doi{10.1017/S0022112007005575}.
\newblock

\bibitem{LiznerAndHornig}
M.~Linzer, D.~F. Hornig, Structure of Shock Fronts in Argon and Nitrogen,
  \emph{Physics of Fluids} 6~(12) (1963) 1661--1668.
  doi:\doi{10.1063/1.1711007}.
\newblock

\bibitem{Camac}
M.~Camac, Argon and nitrogen shock thicknesses, \emph{Proceedings of the Fourth
  International Symposium on Rarefied Gas Dynamics} 1 (1965) 240.
  doi:\doi{10.2514/6.1964-35}.
\newblock

\bibitem{Schmidt1969}
B.~Schmidt, Electron beam density measurements in shock waves in argon,
  \emph{Journal of Fluid Mechanics} 39~(2) (1969) 361--373.
  doi:\doi{10.1017/S0022112069002229}.
\newblock

\bibitem{Garen1974}
W.~Garen, R.~Synofzik, A.~Frohn, Shock Tube for Generating Weak Shock Waves,
  \emph{AIAA Journal} 12~(8) (1974) 1132--1134. doi:\doi{10.2514/3.49425}.
\newblock

\bibitem{Alsemeyer1976}
H.~Alsmeyer, Density profiles in argon and nitrogen shock waves measured by the
  absorption of an electron beam, \emph{Journal of Fluid Mechanics} 74~(3)
  (1976) 497--513. doi:\doi{10.1017/S0022112076001912}.
\newblock

\bibitem{Paolucci2018}
S.~Paolucci, C.~Paolucci, A second-order continuum theory of fluids,
  \emph{Journal of Fluid Mechanics} 846~(22) (2018) 686–710.
  doi:\doi{10.1017/jfm.2018.291}.
\newblock

\bibitem{Chapman1970}
S.~Chapman, T.~G. Cowling, \emph{The Mathematical Theory of Non-Uniform Gases},
  Cambridge University Press, 1970.

\bibitem{Grad1949}
H.~Grad, On the kinetic theory of rarefied gases, \emph{Communications on Pure
  and Applied Mathematics} 2~(4) (1949) 331–407.
  doi:\doi{10.1002/cpa.3160020403}.
\newblock

\bibitem{Grad1952}
H.~Grad, The profile of a steady plane shock wave, \emph{Communications on Pure
  and Applied Mathematics} 5~(3) (1952) 257–300.
  doi:\doi{10.1002/cpa.3160050304}.
\newblock

\bibitem{Agarwal2001}
R.~K. Agarwal, K.-Y. Yun, R.~Balakrishnan, Beyond Navier--Stokes: Burnett
  equations for flows in the continuum transition regime, \emph{Physics of
  Fluids} 13 (2001) 3061–3085. doi:\doi{10.1063/1.1397256}.
\newblock

\bibitem{Jin2001}
S.~Jin, M.~Slemrod, Regularization of the Burnett equations via relaxation,
  \emph{Journal of Statistical Physics} 103~(5-6) (2001) 1009–1033.
  doi:\doi{10.1023/A:1010365123288}.
\newblock

\bibitem{RegBurett2003}
H.~Struchtrup, M.~Torrilhin, Regularization of Grad’s 13 moment equations:
  derivation and linear analysis, \emph{Physics of Fluids} 15~(9) (2003)
  2668–2680. doi:\doi{10.1063/1.1597472}.
\newblock

\bibitem{Woods1993}
L.~C. Woods, \emph{An Introduction to the Kinetic Theory of Gases and
  Magnetoplasmas}, Oxford University Press, 1993.

\bibitem{Bird1963}
G.~A. Bird, Approach to translational equilibrium in a rigid sphere gas,
  \emph{Physics of Fluids} 6 (1963) 1518. doi:\doi{10.1063/1.1710976}.
\newblock

\bibitem{Bird1970}
G.~A. Bird, Direct simulation of the {Boltzmann} equation, \emph{Physics of
  Fluids} 13 (1970) 2676–2681. doi:\doi{10.1063/1.1692849}.
\newblock

\bibitem{lofthouse2008velocity}
A.~J. Lofthouse, L.~C. Scalabrin, I.~D. Boyd, Velocity slip and temperature
  jump in hypersonic aerothermodynamics, \emph{Journal of Thermophysics and
  Heat Transfer} 22~(1) (2008) 38--49. doi:\doi{10.2514/1.31280}.
\newblock

\bibitem{Sone2007}
Y.~Sone, \emph{Molecular Gas Dynamics: Theory, Techniques, and Applications},
  Springer, 2007.

\bibitem{Hadjiconstantinou2024MolecularMO}
N.~G. Hadjiconstantinou, Molecular Mechanics of Liquid and Gas Slip Flow,
  \emph{Annual Review of Fluid
  Mechanics}doi:\doi{10.1146/annurev-fluid-121021-014808}.
\newblock

\bibitem{Lockerby2004}
D.~A. Lockerby, J.~M. Reese, D.~R. Emerson, R.~W. Barber, Velocity boundary
  condition at solid walls in rarefied gas calculations, \emph{Physical Review
  E} 70 (2004) 017303. doi:\doi{10.1103/PhysRevE.70.017303}.
\newblock

\bibitem{Hadjiconstantinou2003}
N.~G. Hadjiconstantinou, Comment on Cercignani’s second-order slip
  coefficient, \emph{Physics of Fluids} 15~(8) (2003) 2352--2354.
  doi:\doi{10.1063/1.1587155}.
\newblock

\bibitem{Ling_Kurzawski_Templeton_2016}
J.~Ling, A.~Kurzawski, J.~Templeton, Reynolds averaged turbulence modelling
  using deep neural networks with embedded invariance, \emph{Journal of Fluid
  Mechanics} 807 (2016) 155–166. doi:\doi{10.1017/jfm.2016.615}.
\newblock

\bibitem{duraisamy2019turbulence}
K.~Duraisamy, G.~Iaccarino, H.~Xiao, Turbulence modeling in the age of data,
  \emph{Annual Review of Fluid Mechanics} 51 (2019) 357--377.
  doi:\doi{10.1146/annurev-fluid-010518-040547}.
\newblock

\bibitem{duraisamy2021perspectives}
K.~Duraisamy, Perspectives on machine learning-augmented Reynolds-averaged and
  large eddy simulation models of turbulence, \emph{Physical Review Fluids}
  6~(5) (2021) 050504. doi:\doi{10.1103/PhysRevFluids.6.050504}.
\newblock

\bibitem{Brunton_2020}
S.~L. Brunton, B.~R. Noack, P.~Koumoutsakos, Machine Learning for Fluid
  Mechanics, \emph{Annual Review of Fluid Mechanics} 52~(1) (2020) 477–508.
  doi:\doi{10.1146/annurev-fluid-010719-060214}.
\newblock

\bibitem{RAISSI2019686}
M.~Raissi, P.~Perdikaris, G.~Karniadakis, Physics-informed neural networks: A
  deep learning framework for solving forward and inverse problems involving
  nonlinear partial differential equations, \emph{Journal of Computational
  Physics} 378 (2019) 686--707.
  doi:\doi{https://doi.org/10.1016/j.jcp.2018.10.045}.
\newblock

\bibitem{KarniadakisPINNS}
G.~Karniadakis, Y.~Kevrekidis, L.~Lu, P.~Perdikaris, S.~Wang, L.~Yang,
  Physics-informed machine learning, \emph{Nature Reviews Physics} 3 (2021)
  422--440. doi:\doi{10.1038/s42254-021-00314-5}.
\newblock

\bibitem{cai2021physicsinformedneuralnetworkspinns}
S.~Cai, Z.~Mao, Z.~Wang, M.~Yin, G.~E. Karniadakis, Physics-informed neural
  networks (PINNs) for fluid mechanics: A review, \emph{Acta Mechanica Sinica}
  37 (2021) 1727--1738. doi:\doi{10.1007/s10409-021-01148-1}.
\newblock

\bibitem{Mao_2021}
Z.~Mao, L.~Lu, O.~Marxen, T.~A. Zaki, G.~E. Karniadakis, DeepM\&Mnet for
  hypersonics: Predicting the coupled flow and finite-rate chemistry behind a
  normal shock using neural-network approximation of operators, \emph{Journal
  of Computational Physics} 447 (2021) 110698.
  doi:\doi{10.1016/j.jcp.2021.110698}.
\newblock

\bibitem{sirignano2020dpm}
J.~Sirignano, J.~F. MacArt, J.~B. Freund, DPM: A deep learning PDE augmentation
  method with application to large-eddy simulation, \emph{Journal of
  Computational Physics} 423 (2020) 109811.
  doi:\doi{10.1016/j.jcp.2020.109811}.
\newblock

\bibitem{Zanardi_2023}
I.~Zanardi, S.~Venturi, M.~Panesi, Adaptive physics-informed neural operator
  for coarse-grained non-equilibrium flows, \emph{Scientific Reports} 13~(1).
  doi:\doi{10.1038/s41598-023-41039-y}.
\newblock

\bibitem{macart2021embedded}
J.~F. MacArt, J.~Sirignano, J.~B. Freund, Embedded training of neural-network
  subgrid-scale turbulence models, \emph{Physical Review Fluids} 6~(5) (2021)
  050502. doi:\doi{10.1103/PhysRevFluids.6.050502}.
\newblock

\bibitem{sirignano2023bluff}
J.~Sirignano, J.~F. MacArt, Deep Learning Closure Models for Large-Eddy
  Simulation of Flows Around Bluff Bodies, \emph{Journal of Fluid Mechanics}
  966 (2023) A26. doi:\doi{10.1017/jfm.2023.446}.
\newblock

\bibitem{sirignano2021pde}
J.~Sirignano, J.~F. MacArt, K.~Spiliopoulos, PDE-constrained models with neural
  network terms: Optimization and global convergence, \emph{Journal of
  Computational Physics} 481 (2023) 112016.
  doi:\doi{10.1016/j.jcp.2023.112016}.
\newblock

\bibitem{kakka2025neuralnetworkaugmentededdyviscosity}
P.~Kakka, J.~F. MacArt, Neural network-augmented eddy viscosity closures for
  turbulent premixed jet flames, \emph{Combustion and Flame} 278 (2025) 114241.
  doi:\doi{10.1016/j.combustflame.2025.114241}.
\newblock

\bibitem{nair2023deep}
A.~S. Nair, J.~Sirignano, M.~Panesi, J.~F. MacArt, Deep Learning Closure of the
  Navier--Stokes Equations for Transition-Continuum Flows, \emph{AIAA Journal}
  61~(12) (2023) 5484--5497. doi:\doi{10.2514/1.J062935}.
\newblock

\bibitem{kryger2024optimizationsecondordertransportmodels}
M.~Kryger, J.~F. MacArt, \href{https://arxiv.org/abs/2411.13515}{Optimization
  of Second-Order Transport Models for Transition-Continuum Flows}, \emph{AIAA
  Journal} accepted, in press.

\bibitem{maxwell1879vii}
J.~C. Maxwell, VII. On stresses in rarified gases arising from inequalities of
  temperature, \emph{Philosophical Transactions of the Royal Society of London}
  170 (1879) 231--256. doi:\doi{10.1098/rstl.1879.0067}.
\newblock

\bibitem{Sharipov2011}
F.~Sharipov, Data on the Velocity Slip and Temperature Jump on a Gas-Solid
  Interface, \emph{Journal of Physical and Chemical Reference Data} 40~(2)
  (2011) 023101. doi:\doi{10.1063/1.3580290}.
\newblock

\bibitem{PhysRevE.81.056314}
M.~A. Solovchuk, T.~W.~H. Sheu, Prediction of shock structure using the bimodal
  distribution function, \emph{Physical Review E} 81 (2010) 056314.
  doi:\doi{10.1103/PhysRevE.81.056314}.
\newblock

\bibitem{GRAUR201187}
I.~Graur, A.~P. Polikarpov, F.~Sharipov, Numerical modeling of rarefied gas
  flow through a slit into vacuum based on the kinetic equation,
  \emph{Computers \& Fluids} 49~(1) (2011) 87--92.
  doi:\doi{https://doi.org/10.1016/j.compfluid.2011.05.001}.
\newblock

\bibitem{karniadakis2005microflows}
G.~E. Karniadakis, A.~Beskok, N.~R. Aluru, \emph{Microflows and Nanoflows:
  Fundamentals and Simulation}, Springer Science \& Business Media, 2005.

\bibitem{sone2007molecular}
Y.~Sone, \emph{Molecular Gas Dynamics: Theory, Techniques, and Applications},
  Birkhäuser, 2007.

\bibitem{GavasaaneEtAll}
A.~Gavasane, S.~Sachdev, B.~Mittal, U.~Bhandarkar, A.~Agrawal, A Critical
  Assessment of the Maxwell Slip Boundary Condition for Rarified Wall Bounded
  Flows, \emph{International Journal of Micro-Nano Scale Transport} 2 (2011)
  109. doi:\doi{10.1260/1759-3093.2.2-3.109}.
\newblock

\bibitem{boyd1995predicting}
I.~D. Boyd, G.~Chen, Predicting failure of the continuum fluid equations in
  transitional hypersonic flows, \emph{Physics of Fluids} 7~(1) (1995)
  210--219. doi:\doi{10.1063/1.868720}.
\newblock

\bibitem{sharipov2011data}
F.~Sharipov, Data on the velocity slip and temperature jump on a gas-solid
  interface, \emph{Journal of Physical and Chemical Reference Data} 29~(2)
  (2011) 021024. doi:\doi{10.1063/1.3580290}.
\newblock

\bibitem{lele1991compact}
S.~Lele, Compact finite difference schemes with spectral-like resolution,
  \emph{Journal of Computational Physics} 103 (1991) 16--42.
  doi:\doi{10.1016/0021-9991(92)90324-R}.
\newblock

\bibitem{MACCORMACK1989135}
R.~W. MacCormack, G.~V. Candler, The solution of the Navier-Stokes equations
  using Gauss-Seidel line relaxation, \emph{Computers \& Fluids} 17~(1) (1989)
  135--150. doi:\doi{10.1016/0045-7930(89)90012-1}.
\newblock

\bibitem{powers2004necessity}
J.~M. Powers, On the necessity of positive semi-definite conductivity and
  Onsager reciprocity in modeling heat conduction in anisotropic media,
  \emph{Journal of Heat Transfer} 126~(5) (2004) 670--675.
  doi:\doi{10.1115/1.1798913}.
\newblock

\bibitem{gurtin2010mechanics}
M.~E. Gurtin, E.~Fried, L.~Anand, \emph{The Mechanics and Thermodynamics of
  Continua}, Cambridge University Press, 2010.

\bibitem{nair2023entropy}
A.~S. Nair, J.~Sirignano, M.~Panesi, J.~F. MacArt, Entropy-stable Deep Learning
  for Navier--Stokes Predictions of Transitional-regime Flows, in: AIAA SciTech
  Forum, 2023, p. 1796.
\newblock

\bibitem{powers2016combustion}
J.~M. Powers, \emph{Combustion Thermodynamics and Dynamics}, Cambridge
  University Press, 2016.

\bibitem{lu2021learning}
L.~Lu, P.~Jin, G.~Pang, Z.~Zhang, G.~E. Karniadakis, Learning nonlinear
  operators via {D}eep{O}{N}et based on the universal approximation theorem of
  operators, \emph{Nature Machine Intelligence} 3~(3) (2021) 218--229.
  doi:\doi{10.1038/s42256-021-00302-5}.
\newblock

\bibitem{plimpton2019direct}
S.~J. Plimpton, S.~Moore, A.~Borner, A.~Stagg, T.~Koehler, J.~Torczynski,
  M.~Gallis, Direct simulation Monte Carlo on petaflop supercomputers and
  beyond, \emph{Physics of Fluids} 31~(8) (2019) 086101.
  doi:\doi{10.1063/1.5108534}.
\newblock

\bibitem{saad2003iterative}
Y.~Saad, \emph{Iterative Methods for Sparse Linear Systems}, Society for
  Industrial and Applied Mathematics, 2003.

\bibitem{kingma2014adam}
D.~P. Kingma, J.~Ba, Adam: A method for stochastic optimization, \emph{arXiv
  preprint arXiv:1412.6980}.

\bibitem{liu2024adjoint}
X.~Liu, J.~F. MacArt, Adjoint-based machine learning for active flow control,
  \emph{Physical Review Fluids} 9 (2024) 013901.
  doi:\doi{10.1103/PhysRevFluids.9.013901}.
\newblock

\bibitem{hickling2024large}
T.~Hickling, J.~Sirignano, J.~F. MacArt, Large Eddy Simulation of Airfoil Flows
  Using Adjoint-Trained Deep Learning Closure Models, in: AIAA SciTech Forum,
  2024, p. 0296.
\newblock

\bibitem{paszke2019pytorchimperativestylehighperformance}
A.~Paszke, S.~Gross, F.~Massa, A.~Lerer, J.~Bradbury, G.~Chanan, T.~Killeen,
  Z.~Lin, N.~Gimelshein, L.~Antiga, A.~Desmaison, A.~Köpf, E.~Yang, Z.~DeVito,
  M.~Raison, A.~Tejani, S.~Chilamkurthy, B.~Steiner, L.~Fang, J.~Bai,
  S.~Chintala, PyTorch: An Imperative Style, High-Performance Deep Learning
  Library (2019).
\newblock \href {http://arxiv.org/abs/1912.01703} {\path{arXiv:1912.01703}}.

\bibitem{9485033}
C.~R. Trott, D.~Lebrun-Grandié, D.~Arndt, J.~Ciesko, V.~Dang, N.~Ellingwood,
  R.~Gayatri, E.~Harvey, D.~S. Hollman, D.~Ibanez, N.~Liber, J.~Madsen,
  J.~Miles, D.~Poliakoff, A.~Powell, S.~Rajamanickam, M.~Simberg,
  D.~Sunderland, B.~Turcksin, J.~Wilke, Kokkos 3: Programming Model Extensions
  for the Exascale Era, \emph{IEEE Transactions on Parallel and Distributed
  Systems} 33~(4) (2022) 805--817. doi:\doi{10.1109/TPDS.2021.3097283}.
\newblock

\bibitem{dataset}
A.~S. Nair, J.~F. MacArt, Code and Dataset Accompanying ``Physics-Based Machine
  Learning Closures and Wall Models for Hypersonic Transition–Continuum
  Boundary Layer Predictions'', Available:
  \url{https://doi.org/10.5281/zenodo.15678863}, {June 2025}.
\newblock

\end{thebibliography}
\end{document}